\documentstyle[prl,multicol,aps,amsmath]{revtex}

\newcommand{\extraspace}{\addtolength{\abovedisplayskip}{2mm}
                        \addtolength{\belowdisplayskip}{2mm}
                        \addtolength{\abovedisplayshortskip}{2mm}
                        \addtolength{\belowdisplayshortskip}{2mm}}
\newcommand{\be}{\begin{equation}\extraspace}
\newcommand{\ee}{\end{equation}}
\newcommand{\bea}{\begin{eqnarray}\extraspace}
\newcommand{\eea}{\end{eqnarray}}

\def\top#1{\vskip #1\begin{picture}(290,80)(80,500)\thinlines
\put(65,500){\line( 1, 0){255}}\put(320,500){\line( 0, 1){5}}
\end{picture}}
\def\bottom#1{\vskip #1\begin{picture}(290,80)(80,500)\thinlines
\put(330,500){\line( 1, 0){255}}\put(330,500){\line( 0, -1){5}}
\end{picture}}
\input{epsf.tex}
\def\inseps#1#2{\def\epsfsize##1##2{#2##1} \centerline{\epsfbox{#1}}}

\renewcommand{\narrowtext}{\begin{multicols}{2}
\global\columnwidth20.5pc}
\renewcommand{\widetext}{\end{multicols}
\global\columnwidth42.5pc}
\newfont{\BBB}{msbm10 scaled\magstephalf}

\newcommand{\BBN}{\mbox{\BBB N}}

\DeclareMathAlphabet{\mathbb}{U}{bbold}{m}{n}
\newcommand{\id}{\mathbb 1}

\newcommand{\wtPsi}{\widetilde{\Psi}}

\newcommand{\nonu}{\nonumber \\[2mm]}

\begin{document}
\draft

\title{Non-Abelian spin-singlet quantum Hall states: \\
wave functions and quasihole state counting}
\author{Eddy Ardonne$^1$, N. Read$^2$, Edward Rezayi$^3$, and
Kareljan Schoutens$^1$}
\address{
$^1$Institute for Theoretical Physics, Valckenierstraat 65, 1018
XE Amsterdam, The Netherlands \\ $^2$Department of Physics, Yale
University, P.O. Box 208120, New Haven, CT 06520-8120\\
$^3$Department of Physics, CSU-Los Angeles, Los Angeles, CA 90032}
\date{April 2001}
\maketitle
\begin{abstract}
We investigate a class of non-Abelian spin-singlet (NASS) quantum
Hall phases, proposed previously. The trial ground and quasihole
excited states are exact eigenstates of certain $k+1$-body
interaction Hamiltonians. The $k=1$ cases are the familiar
Halperin Abelian spin-singlet states. We present closed-form
expressions for the many-body wave functions of the ground states,
which for $k>1$ were previously defined only in terms of
correlators in specific conformal field theories. The states
contain clusters of $k$ electrons, each cluster having either all
spins up, or all spins down. The ground states are non-degenerate,
while the quasihole excitations over these states show
characteristic degeneracies, which give rise to non-Abelian braid
statistics. Using conformal field theory methods, we derive
counting rules that determine the degeneracies in a spherical
geometry. The results are checked against explicit numerical
diagonalization studies for small numbers of particles on the
sphere.

\end{abstract}


\vskip -4mm

%
%
%

\narrowtext

\section{Introduction and Summary}

The observation \cite{willett,pan1,read} of a quantum Hall (QH)
state at an even-denominator filling factor, $\nu=5/2$, stimulated
the development of trial wave functions outside the usual
hierarchy (or later, composite fermion) approach, which generates
only odd-denominator fractions. The $5/2$ state is interpreted as
half-filling of the first excited Landau level (LL), the lowest
one being filled with electrons of both spins, and can be mapped
to half-filling of the lowest LL, with a suitable Hamiltonian.
There are now strong indications that this state is spin-polarized
\cite{RH,pan2}, and described \cite{RH} by the paired ``Pfaffian''
state of Moore and Read (MR) \cite{mr1}, which has filling factor
$1/2$. This state was originally proposed as an example that
manifests non-Abelian braid statistics of its quasiparticle
excitations \cite{mr1}. Generalizations exist in which the
particles are ``clustered'' in $k$-plets ($k=1$, $2$, $3$,
\ldots), but still spin-polarized \cite{rr2}. In these states, the
non-Abelian statistics are associated with parafermion conformal
field theories (CFTs).

It is well known that, despite the presence of a strong magnetic
field, spin-singlet QH states are sometimes favored over their
spin-polarized counterparts. The possibility to manipulate the
Zeeman splitting by the application hydrostatic pressure and by
tilting the magnetic field has opened the possibility of
systematic studies of transitions between spin-polarized and
non-polarized QH states (see, e.g., \cite{spintr}). In this light,
states analogous to the spin-polarized clustered QH states of
\cite{rr2}, but with a spin-singlet structure, have been
constructed \cite{as1}. In \cite{as1}, trial wave functions for
these non-Abelian spin-singlet (NASS) states have been written in
terms of correlators in a CFT describing parafermions associated
to the Lie algebra SU($3$). We remark that one may consider an
alternative series of NASS states, whose algebraic structure is
related to SO($5$) rather than SU($3$). The simplest state of this
type is a paired spin-singlet state that exhibits a separation of
spin and charge in the quasihole excitation spectrum \cite{alls}.
The SO($5$)-based NASS states will not be discussed in this paper.

In the present paper, we study in detail some of the properties of
the NASS states, paying special attention to the case $k=2$. We
give explicit closed form expressions for the ground state wave
functions, and study the degeneracies of their quasihole
excitations. The degeneracies of states with fixed spins and fixed
well-separated positions of the quasiparticles are the origin of
the non-Abelian braid statistics.

We first strengthen the case for the existence of the
incompressible phases of matter with the universal properties of
the states of Ardonne and Schoutens (AS) \cite{as1}, by showing
that their trial wave functions for the ground state and for
states with quasiholes are exact zero-energy eigenstates of
certain $k+1$-body interaction Hamiltonians for particles in a
single LL, in a similar way as the spin-polarized cases
\cite{rr2}. The explicit closed-form wave functions for the ground
states are obtained. In the study of the quasihole degeneracies,
we then follow two complementary approaches. The first is an
analytical path, which relies heavily on the formal structure of
the associated parafermion CFT, and on the analogy with earlier
studies for spin-polarized non-Abelian QH states
\cite{nayak,rr1,gr1}. While at present we lack explicit
expressions for the many-body wave functions describing the
quasiholes, we have enough control to derive explicit counting
formulas for the degeneracies, for $k=2$, for particles on a
sphere. The second approach is a numerical study of the $k+1$-body
Hamiltonian for the case $k=2$, on the sphere. The numbers of
zero-energy states for each number of electrons $N$ and of
quasiholes $n$ considered are in exact agreement with the
analytical derivation. In addition, we study the excitation
spectrum of the same Hamiltonian, and compare the ground state
with that of electrons interacting via the lowest LL Coulomb
interaction.

A highlight of the analytical approach in this paper comes in the
derivation of the total degeneracies of quasihole states. In the
CFT set up (which will be described in more detail in Section
\ref{cft}) the QH states are described as conformal blocks of
``particle'' and ``quasihole'' operators. The particle operator
factorizes as a product of a vertex operator and a parafermion
field, and the quasihole operator is the product of a vertex
operator and a so-called spin field of the parafermion CFT. The
non-trivial fusion rules of these spin fields cause a degeneracy
of the ground states in the presence of quasiholes at fixed
positions and spins. There is also further degeneracy associated
with the positions and spins of the quasiholes, which is finite on
the sphere. The two contributions need to be combined in the right
way. It turns out, as in earlier cases, that the various states
which stem from the non-trivial fusion rules have a different
spatial degeneracy. Therefore, we can not just multiply the two
degeneracy factors, but we need to break up the degeneracies due
to the fusion rules. To accomplish this task, and arrive at the
final counting formula, we analyze ``truncated chiral spectra'' in
the SU($3$) parafermion CFT, using the methods of \cite{sc1}.

This paper is organized as follows. In Section \ref{cft}, we
explain in which way CFT is used to describe QH states, and review
the NASS states as correlators. In Section \ref{sec:Ham}, we
introduce the $k+1$-body interaction, and show that the AS
correlators give zero-energy ground states. In Section \ref{gswf}
we give the explicit ground state wave functions for the NASS
states, and discuss their spin-singlet properties. Section
\ref{qhcor} describes the correlators which give the states with
quasiholes present. The derivation of the counting formula is
done in Sections \ref{fusions} to \ref{cfor}, using the method
which is outlined above, with Eq.\ (\ref{genfor}) as the final
outcome. Explicit results of this formula for the degeneracy of
the ground states in the presence of quasiholes are given in the
same Section for several $N$ (the number of electrons) and $n$
(the number of quasiholes). In Section \ref{numerics} we present
numerical diagonalization studies on a sphere, finding full
agreement with the analytical expression obtained in Section
\ref{cfor}, and compare the states with the ground state of the
Coulomb interaction.

\section{QHE-CFT correspondence}
\label{cft}

In the QHE, following Laughlin \cite{laugh}, trial wave functions
have long been used as paradigms that represent an entire phase of
incompressible behavior. This notion was reviewed in Ref.\
\cite{rr2}, so in this paper we will concentrate on the properties
of trial states and their position-space wave functions. As
explained by MR \cite{mr1}, many QH trial wave functions of the
$2+1$ dimensional system can be obtained as conformal blocks
(i.e., chiral parts of correlation functions or ``correlators'')
in a suitable chiral CFT in $2$ Euclidean dimensions, as we will
briefly explain.

For particles with complex coordinates $z_j=x_j+iy_j$, $j=1$,
\ldots, $N$, for $N$ particles, we will use reduced wave functions
$\wtPsi(z)$, and neglect spin temporarily. For the lowest Landau
level (LLL), the reduced wave function must be holomorphic in the
$z_i$'s. For particles in the plane, the full wave function
$\Psi(z)$ is recovered by multiplying $\wtPsi(z)$ by
$\exp{(-\sum_i |z_i|^2/4})$; we have set the magnetic length to
$1$. For particles on the sphere \cite{hald}, the coordinate $z_i$
refers to stereographic projection, and the full wave function is
recovered by multiplying by
$\prod_i(1+|z_i|^2/4R^2)^{-(1+N_\phi/2)}$, where $N_\phi$ is the
total number of magnetic flux quanta through the sphere
\cite{rr1,rr2}. In the latter case, the reduced wave function
$\wtPsi(z)$ must be a polynomial of degree no higher than $N_\phi$
in each $z_i$, so that the $z$ component of angular momentum of
each particle lies between $N_\phi/2$ and $-N_\phi/2$. Note that
we use the term ``particles'' for the underlying particles, which
are either charged bosons or charged fermions (electrons), because
it is convenient to consider cases of either statistics together.

The simplest example of a state with a uniform density (a state of
zero total angular momentum on the sphere \cite{hald}) is the
Laughlin wave function \cite{laugh}:
\begin{equation}
\wtPsi_{\rm L}^M(z_1,\ldots,z_N) = \prod_{i<j}(z_i-z_j)^M \ ,
\label{eq:lau}
\end{equation}
for a fixed integer $M$. The filling factor can be defined for a
sequence of states as $\nu=\lim_{N\to\infty}N/N_\phi$, where
$N_\phi$ is identified with the largest power of any $z_i$ in the
state. For the Laughlin state it is $\nu=\frac{1}{M}$. Note that
this function is antisymmetric (describes fermions) for $M$ odd,
and symmetric (describes bosons) for $M$ even.

The Laughlin wave function can be obtained as
\begin{eqnarray}
&& \wtPsi_{\rm L}^M (z_1,\ldots,z_N) = \nonu \label{lacor}
&& \lim_{z_\infty \to \infty} (z_\infty)^a
\langle V_e(z_1) \ldots V_e(z_N)
e^{- i \sqrt{M} N \varphi}(z_\infty) \rangle \ ,
\end{eqnarray}
with $V_e(z)=\exp (i \sqrt{M} \varphi)$ a chiral vertex operator
in the $c=1$ chiral CFT describing a single scalar field $\varphi$
compactified on a radius $R^2=M$. The operator $e^{- i \sqrt{M} N
\varphi}(z_\infty)$ brings in a positive background charge, which
guarantees the overall neutrality of the system. The constant $a$
must be chosen in such a way that the effect of the background
charge does not go to zero in the limit $z_\infty \to \infty$; in
Eq.\ (\ref{lacor}) we need $a=MN^2$. This procedure is simpler for
our purposes than the uniform background charge used in MR
\cite{mr1}, though the latter has the additional feature of
reproducing the factors in the unreduced wave function.

Other, more complicated, QH states can be constructed by invoking
more complicated CFTs. The CFT framework guarantees that a number
of consistency requirements for such states are met \cite{mr1}.
The trial wave functions become more meaningful, and the
corresponding phase really exists, when there is a (local)
Hamiltonian for which the trial state is the nondegenerate ground
state, and the excitation spectrum (for the same $N_\phi$ as the
ground state) has a gap in the thermodynamic limit. Short range
2-body interaction Hamiltonians with these properties for the
Laughlin state were found by Haldane \cite{hald}, and 3-body
interactions for which the MR state is an exact zero energy
eigenstate were found beginning from the work of Ref.\ \cite{gww}.
Read and Rezayi (RR) discovered \cite{rr2} that these
constructions are the first two cases in an infinite sequence, and
found the parafermion states as the exact zero energy eigenstates
of $k+1$-body interactions for all $k$. Here we will show
similarly that, in the case of particles with spin, the NASS
states of Ref.\ \cite{as1} are exact eigenstates of zero energy
for $k+1$ body spin-independent interactions, with the Halperin
state and the corresponding 2-body interaction \cite{hald} as the
only case known previously. First, we recall the construction in
Ref.\ \cite{as1}, then in the following Section we establish that
the wave function defined by a correlator here is a zero-energy
eigenstate of a $k+1$-body interaction Hamiltonian.

The NASS states proposed in \cite{as1} can be viewed as
non-Abelian generalizations of the Abelian spin-singlet Halperin
states labeled as $(m+1,m+1,m)$ (see Eq.\ (\ref{wf3}) below), or
alternatively, as spin-singlet analogs of the spin-polarized
``clustered'' or ``parafermion'' states of RR \cite{rr2}. The
filling fraction of the NASS states is $\nu = \frac{2k}{2kM+3}$,
with $M$ an integer ($M$ is odd when the particles are fermions,
even when they are bosons). The wave functions of these states are
constructed as conformal blocks in basically the same way as was
done above for the Laughlin state. In the basic case $M=0$, the
component of the ground state with any set of $N/2$ of the
particles having spin up, and the remainder spin down, is defined
(up to phases that may be needed when reconstructing the full
state from these components) as the correlator \cite{as1}
\begin{eqnarray}
&& \wtPsi_{\rm NASS}^{k,M=0} = \lim_{z_\infty \to \infty}
(z_\infty)^\frac{3N^2}{2k} \, \langle
\exp(i\frac{N}{2\sqrt{k}}(\alpha_2-\alpha_1)\cdot\vec{\varphi})
(z_\infty) \nonu && \times B_{\alpha_1}(z_1^\uparrow)\ldots
B_{\alpha_1}(z_{N/2}^\uparrow) B_{-\alpha_2}(z_1^\downarrow)
\ldots B_{-\alpha_2}(z_{N/2}^\downarrow) \rangle \ . \label{wf0}
\end{eqnarray}
In this equation the ``particle operators'' (to avoid confusion,
we emphasize that this means the operators that correspond to the
particles in the CFT, not the operators that create actual
particles in the $2+1$-dimensional system) are current operators
$B_\alpha(z)$ in an SU($3)_k$ (i.e., level $k$) Wess-Zumino-Witten
CFT. These currents can be written in terms of two bosons
$\vec{\varphi} = (\varphi_1,\varphi_2)$ and a Gepner parafermion
$\psi_\alpha$ associated to SU($3)_k/[$U($1$)]$^2$ \cite{gep1}.
The currents are labeled by the corresponding roots $\alpha$ of
SU($3$)
\begin{equation}
B_\alpha(z) = \psi_{\alpha}
\exp(i \alpha \cdot \vec{\varphi} / \sqrt{k})(z) \ .
\end{equation}
The roots are given by $\alpha_1= (\sqrt{2},0)$,
$\alpha_2=(-\sqrt{2}/2,\sqrt{6}/2)$. We note that these two roots
form an SU($2$) doublet under an SU($2$) subalgebra of SU($3$);
the embedding of the subalgebra is isomorphic to that given in
terms of $3\times 3$ Hermitian matrices [generators of SU($3$)] by
the $2\times 2$ Hermitian blocks at the upper left corner. For
such an embedding, there is also a U($1$) subalgebra [generated by
``hypercharge'' diag($1,1,-2$); the particles carry hypercharge 1]
that commutes with the SU($2$) and corresponds to the particle
number. Note that a (hyper-)charge at infinity is again needed for
neutrality.

Working out the vertex-operator part of this correlator, we arrive
at the following factorized form of the NASS state (after
multiplication by an additional Laughlin factor to obtain general
$M$\footnote{We can also obtain the Laughlin factor by using the
particle operators (\protect\ref{elup}) and (\protect\ref{eldown})
in the correlator, together with a suitably adjusted background
charge.} )
\begin{eqnarray}
&& \wtPsi_{\rm NASS}^{k,M}(z_1^\uparrow,\ldots,z_{N/2}^\uparrow;
   z_1^\downarrow,\ldots,z_{N/2}^\downarrow)=  \nonu
&& \langle \psi_{\alpha_1}(z_1^\uparrow) \ldots
\psi_{\alpha_1}(z_{N/2}^\uparrow) \psi_{-\alpha_2}(z_1^\downarrow)
\ldots \psi_{-\alpha_2} (z_{N/2}^\downarrow) \rangle \nonu &&
\times \left[ \wtPsi_{\rm
H}^{(2,2,1)}(z_i^\uparrow;z_j^\downarrow) \right]^{1/k} \,
\wtPsi^M_{\rm L}(z_i^\uparrow;z_j^\downarrow) \ . \label{wf1}
\end{eqnarray}
The wave function $\widetilde{\Psi}_{\rm H}^{(2,2,1)}$ is one of
the Halperin wave functions \cite{halp1}
\begin{eqnarray}
&& \widetilde{\Psi}^{(m',m',m)}_{\rm H}
(z_1^\uparrow,\ldots,z_{N/2}^\uparrow;
z_1^\downarrow,\ldots,z_{N/2}^\downarrow) = \nonu && \label{wf3}
\prod_{i<j} (z_i^\uparrow-z_j^\uparrow)^{m'} \prod_{i<j}
(z_i^\downarrow-z_j^\downarrow)^{m'} \prod_{i,j}
(z_i^\uparrow-z_j^\downarrow)^m \ .
\end{eqnarray}
The latter give rise to spin-singlet states whenever $m'=m+1$
\cite{hald}. The wave function Eq.\ (\ref{wf1}) contains a term
which is a correlator of parafermions, the explicit form of which
will be found below.

We also mention here that the CFT construction implies that the
number of sectors of edge states (representations of the chiral
algebra), and the number of ground states (conformal blocks with
$N$ particle operators inserted) on the torus for $N$ divisible by
$2k$, are both given by $(k+1)(k+2)(2kM+3)/6$, which is an
integer. For $M=0$, this coincides with the numbers for SU$(3)_k$
current algebra. The filling factor is $P/Q=2k/(2kM+3)$, so if $P$
and $Q$ are defined as being coprime, then the denominator
$Q=2kM+3$ unless $2k$ and $3$ have a common factor, that is unless
$k$ is divisible by 3, in which case we have $Q=(2kM+3)/3$. The
number of ground states on the torus is then always divisible by
the denominator $Q$ of the filling factor, as it should be. We
also expect that for some $k$ values there are ground states on
the torus for other $N$ values, as for the MR states \cite{rg} and
RR states.

\section{Solution of $k+1$-body Hamiltonian}
\label{sec:Ham}

It is known \cite{hald} that the Abelian Halperin spin-singlet
state is the unique (on the sphere) exact zero-energy eigenstate
of a two-body interaction Hamiltonian. Other than the trivial case
(1,1,0), which is two filled Landau levels, where the Hamiltonian
in question is zero, the simplest case is $m=1$ in Eq.\
(\ref{wf3}), the (2,2,1) state, which corresponds to $M=0$ in Eq.\
(\ref{wf0}) or (\ref{wf1}). In the latter case this Hamiltonian is
simply a repulsive $\delta$-function interaction between any two
particles. As in Ref.\ \cite{rr1,rr2}, it is simplest to start by
generalizing this $M=0$ case. Because higher $M$ values are
obtained by multiplying by additional Laughlin factors, the
Hamiltonians for $M=0$ can be straightforwardly extended to $M>0$
by extending the range of the $k+1$-body part, and adding 2-body
interactions as needed, which have the effect of requiring
zero-energy states to contain the Laughlin factors. We will not
detail this here, however, see Sec.\ \ref{numerics} below.

The natural choice of Hamiltonian for $M=0$ is to consider the
$k+1$-body $\delta$-function as in Ref.\ \cite{rr2}, but here for
particles with spin. The Hamiltonian (acting within the LLL) is
\begin{eqnarray}
H &=& V\!\!\!\sum_{i_1<i_2<\cdots<i_{k+1}}
\delta^{2}(z_{i_1}-z_{i_2}) \nonu &&
\times \delta^{2}(z_{i_2}-z_{i_3})
\cdots\delta^2(z_{i_{k}}-z_{i_{k+1}}), \end{eqnarray} with $V>0$.
Note that here we have reverted to labeling the particles
independently of their spin. For this Hamiltonian, a state is a
zero-energy eigenstate if it vanishes whenever any $k+1$ particles
coincide; for this to be satisfied for some nontrivial states, the
particles must be bosons.

We will now show that the correlator as in Eq.\ (\ref{wf0}) is
such a zero-energy eigenstate. The argument we give here makes
direct use of the current algebra satisfied by the currents, and
also sheds new light on the previous spinless case of RR, where a
slightly different argument was used. Without loss of generality,
we can consider letting the first $k+1$ particles, of either spin,
come to the same point, say $z=0$, that is $z_1^{\sigma_1}$,
$z_2^{\sigma_2}$, \ldots, $z_{k+1}^{\sigma_{k+1}}$ all $\to 0$. In
the standard radial quantization scheme for CFT, we can consider
the current operators as acting in a Hilbert space that is built
starting from a highest weight state that in the present case is
simply the vacuum $|0\rangle$ of radial quantization about the
origin $z=0$. As $z_i^{\sigma_i}$ tend to $0$ one by one, the
resulting operator product expansion (ope) contains no singular
terms. This follows from the standard current-algebra ope's of the
currents, together with the fact that the roots $\alpha_1$ and
$-\alpha_2$ do not sum to either 0 or another root (for
simplicity, let us replace these two roots by the natural notation
$\sigma=\uparrow$ and $\downarrow$, respectively). Indeed the only
nonvanishing term as the $z_i^{\sigma_i}$ tend to zero
sequentially is the operator at $0$ that corresponds to the state
\begin{equation}
B_{\sigma_1,-1}B_{\sigma_2,-1}\cdots B_{\sigma_{k+1},-1}|0\rangle
\end{equation}
in the highest weight representation, and we need to show that
this vanishes for all choices of $\sigma_1$, \ldots,
$\sigma_{k+1}$. Here we have used the modes of the currents,
\begin{equation}
B_a(z)=\sum_n z^{-n-1}B_{a,n},
\end{equation}
which holds for all generators $a$ of SU($3$), not only roots. In
fact, the commutation relations of the affine Lie algebra for
these modes also imply that $B_{\uparrow,-1}$ and
$B_{\downarrow,-1}$ commute, so we need only to prove
\begin{equation}
(B_{\downarrow,-1})^m(B_{\uparrow,-1})^{k+1-m}|0\rangle=0,
\end{equation}
for all $m$ in the range $0\leq m \leq k+1$.

Let us begin by choosing $m=0$. Then we need to show that
\begin{equation}
(B_{\uparrow,-1})^{k+1}|0\rangle=0.
\end{equation}
But this is simply the pure-current-algebra null-vector equation,
which first entered the physics literature in Refs.\
\cite{gepw,fz}. Thus this is satisfied in the irreducible, unitary
vacuum highest weight module of the SU($3$) affine Lie algebra at
level $k$, and there are similar equations for all current
algebras, and for each ``integrable'' highest weight
representation. This already suffices to rederive the RR states,
which are related to SU(2) current algebra \cite{rr2}. The RR
states are in fact the same as those in Eq.\ (\ref{wf0}) above,
but with all spins $\uparrow$, so the only root that appears lies
in an SU($2$) subalgebra (the spin singlet property of the above
states is of course lost when this is done, and the charge at
infinity should be adjusted). Hence, we have very quickly
rederived the fact that the RR states for $M=0$ are the
zero-energy states of the $k+1$-body $\delta$-function
Hamiltonian.

It is straightforward to complete the proof for the NASS states.
Essentially, we use the SU($2$) symmetry under which
$B_{\uparrow,-1}$ and $B_{\downarrow,-1}$ form a doublet, and
notice that the set of states labeled by $m$ forms a highest
weight multiplet under this algebra, of SU($2$) spin $k/2$. (This
is clear from the $2+1$ point of view, where we are looking at
states of $k+1$ spin $1/2$ bosons all at the same point.) Then
since the highest weight vanishes, all the others do also, which
completes the proof.

As in RR, a similar argument also establishes that quasihole
states, written as similar correlators but with spin fields
(primary fields of the SU($3$) current algebra) inserted at the
quasihole positions \cite{as1}, are exact zero-energy eigenstates.
These are considered explicitly in Sec.\ \ref{qhcor} below.
Similar arguments also imply that the zero-energy ground states of
$H$ on the torus are given by correlators for some number $N$ of
the above fields inserted, with the number of such ground states
(for $N$ divisible by $2k$) already given in Sec.\ \ref{cft}.

\section{Ground state wave function}
\label{gswf}

Based on the results of the previous Section, we expect the
structure of the trial wave functions---that is, of the chiral
correlators (\ref{wf1})---of the NASS states to be similar to that
of the RR states, and also to generalize the Halperin (2,2,1)
state. The RR wave functions were constructed by dividing the
(same-spin) particles into clusters of $k$, writing down a product
of factors for each pair of clusters, and finally symmetrizing
over all ways of dividing the particles into clusters. Hence in
the case with spin, we guess that we should divide the up
particles into groups of $k$, the downs into groups of $k$ and
then multiply together factors that connect up with up, down with
down, or up with down clusters, and finally ensure that the
function is of the correct permutational symmetry type to yield a
spin-singlet state (in particular, it should be symmetric in the
coordinates of the up particles, and also in those of the downs).
We expect that the up-up and down-down parts of this should
closely resemble the RR wave functions, before the symmetrization;
it was shown in Ref.\ \cite{rr2} that the functions found there
vanish when $k+1$ particles come to the same point, even inside
the sum over permutations that symmetrizes the final function.
These considerations guided the following construction.

Due to the spin-singlet nature of the state, the wave function
will be non-zero only if the number of spin-up and spin-down
particles is the same. Furthermore, there must be an integer
number of clusters, so the total number of particles $N$ must be
divisible by $2k$, and will be written as $N=2kp$, where $p \in
\BBN$. One example was already given in \cite{as1}, namely the
wave function for the case $k=2,M=0$ with the number of particles
equal to $4$ (i.e., $p=1$),
\begin{eqnarray}
&& \widetilde{\Psi}^{k=2,M=0}_{\rm
NASS}(z_1^\uparrow,z_2^\uparrow; z_1^\downarrow,z_2^\downarrow) =
\nonu && \label{wf4} (z_1^\uparrow-z_1^\downarrow)
(z_2^\uparrow-z_2^\downarrow) + (z_1^\uparrow-z_2^\downarrow)
(z_2^\uparrow-z_1^\downarrow) \ .
\end{eqnarray}
This is part of the two-dimensional irreducible representation of
the permutation group on $4$ objects, $S_4$, as can easily be
seen. This is the correct symmetry type to obtain a spin-singlet
state, as we discuss further below.

We will now describe the different factors that enter the NASS
wave functions. Because the only effect of $M$ being non-zero is
to give an overall Laughlin factor, we will assume at first that
$M=0$. First we give the factors that involve particles of the
same spin, say spin up. They are the same as in RR \cite{rr2}. We
will divide the particles into clusters of $k$ in the simplest
way,
\begin{equation} \label{wf5}
(z_1^\uparrow,\ldots,z_k^\uparrow),
(z_{k+1}^\uparrow,\ldots,z_{2k}^\uparrow), \ldots,
(z_{(p-1)k+1}^\uparrow,\ldots,z_{pk}^\uparrow) \ ,
\end{equation}
and the same for the $z^\downarrow$'s. (In a more precise
treatment, we would say that the first $N/2$ particles are spin
up, the remainder spin down.) We write down factors that connect
the $a$th with the $b$th cluster:
\begin{eqnarray}
&& \chi_{a,b}^{z^\uparrow} =
(z_{(a-1)k+1}^\uparrow-z_{(b-1)k+1}^\uparrow)
(z_{(a-1)k+1}^\uparrow-z_{(b-1)k+2}^\uparrow) \nonu && \times
(z_{(a-1)k+2}^\uparrow-z_{(b-1)k+2}^\uparrow)
(z_{(a-1)k+2}^\uparrow-z_{(b-1)k+3}^\uparrow) \nonu && \times
\ldots \times (z_{ak}^\uparrow-z_{bk}^\uparrow)
(z_{ak}^\uparrow-z_{(b-1)k+1}^\uparrow) \ . \label{wf8}
\end{eqnarray}
For $k=1$, we would write
$\chi_{a,b}^{z^\uparrow}=(z_{a}^\uparrow-z_{b}^\uparrow)^2$. The
factors that connect up with down spins are simpler:
\begin{eqnarray}
&& \chi^{z^\uparrow,z^\downarrow}_{a,b} =
(z_{(a-1)k+1}^\uparrow-z_{(b-1)k+1}^\downarrow) \nonu &&
\label{wf6} \times (z_{(a-1)k+2}^\uparrow-z_{(b-1)k+2}^\downarrow)
\ldots (z_{ak}^\uparrow-z_{bk}^\downarrow) \ .
\end{eqnarray}
For $k=1$, the factor would be
$\chi_{a,b}^{z^\uparrow,z^\downarrow}=(z_{a}^\uparrow-z_{b}^\downarrow)$.
We multiply all these factors for all pairs of clusters, up-up,
down-down, or up-down:
\begin{equation} \label{wf11}
\prod_{a<b}^p \chi_{a,b}^{z^\uparrow} \; \;
\prod_{c,d}^p \chi_{c,d}^{z^\uparrow,z^\downarrow} \; \;
\prod_{e<f}^p \chi_{e,f}^{z^\downarrow} \ .
\end{equation}
Notice that for $k=1$, we do obtain the Halperin (2,2,1) wave
function.

To obtain a spin-singlet state when the spatial function is
combined with the spin state (which lies in the tensor product of
$N$ spins $1/2$), some symmetry properties must be satisfied. For
the $M=0$ case, the particles are bosons, hence the full wave
function must be invariant under permutations of spins and
coordinates of any two particles. This can be used to obtain the
correct form of the function from that component in which, say the
first $N/2$ are spin up, the rest spin down, as above, so
knowledge of that component is sufficient. The requirement that
the full wave function be a spin-singlet can be shown to reduce to
the Fock conditions: the component just defined must be symmetric
under permutations of the coordinates of the up particles, and
also of the down particles, and must also obey the Fock cyclic
condition, as given in Ref.\ \cite{hamer} (modified in an obvious
way for the boson case). These three conditions can be shown to
imply that the spatial wave function is of a definite
permutational symmetry type (belongs to a certain irreducible
representation of the permutation group), that corresponds to the
Young diagram with two rows of $N/2$ boxes each. In general, given
a function of arbitrary symmetry, a Young operator can be
constructed that projects it onto a member of the correct
representation (though the result may vanish); this construction
generalizes the familiar symmetrization and antisymmetrization
operations. For the present case, the Young operator is the
following operation, equivalent to summing over the function with
various permutations of its arguments, and some sign changes:
First, antisymmetize in $z_1$, $z_{N/2+1}$; then in $z_2$,
$z_{N/2+2}$; \ldots, $z_{N/2}$, $z_{N}$; then symmetrize in $z_1$,
\ldots, $z_{N/2}$; then finally symmetrize in $z_{N/2+1}$, \ldots,
$z_N$. This clearly satisfies the first two requirements of Fock,
and can be proved to satisfy also the cyclic condition. It remains
to check that it is nonzero, we believe it is. Incidentally, the
application of the Young operator is the analog of symmetrizing
over the down spins in the spatial wave function of the permanent
state (see e.g.\ Ref.\ \cite{rr1}), to which it reduces for the
case of BCS paired wave functions of spin 1/2 bosons (there are
similar statements in the more familiar case of spin-singlet
pairing of spin-1/2 fermions). However, based on the example of
the Halperin ($k=1$) case, we also considered the function defined
as in Eq.\ (\ref{wf11}), and then simply symmetrized over all the
ups and over all the downs. For the Halperin function [which in
fact is already symmetric in Eq.\ (\ref{wf11})], this satisfies
the cyclic condition, as can be seen using the fact that the
(1,1,0) state is a Landau level filled with both spins, plus the
Pauli exclusion principle for fermions. For $k=2$, $3$, we
verified the cyclic condition numerically for several moderate
sizes. Hence, we expect that this simpler form actually works for
all $k$ (as well as for all $N$ divisible by $k$). Apparently,
this procedure and the application of the Young operator give the
same function in the end (up to a normalization).

For $M=0$, our wave function is then:
\begin{equation} \label{wf13}
\widetilde{\Psi}_{\rm NASS}^{k,0} = {\rm Sym}\prod_{a<b}^p
\chi_{a,b}^{z^\uparrow} \; \; \prod_{c,d}^p
\chi_{c,d}^{z^\uparrow,z^\downarrow} \; \; \prod_{e<f}^p
\chi_{e,f}^{z^\downarrow}
\end{equation}
where $\rm Sym$ stands for the symmetrization over the ups and
also over the downs. This function is nonzero, as may be seen by
letting the up coordinates coincide in clusters of $k$ each, and
also the downs, all clusters at different locations, and making
use of the result in RR \cite{rr2} that only one term in the
symmetrization is nonzero in the limit. This term is the Halperin
$(2k,2k,k)$ function for $2p$ particles. To obtain the wave
function for general $M$, we multiply by an overall Laughlin
factor, $\wtPsi_{\rm L}^M$.

We can give a simple proof that our wave function (for $M=0$)
vanishes if any $k+1$ particles, each of either spin, come to the
same point. This works also for the RR wave functions, and is
simpler, though less informative, than the proof in RR \cite{rr2}.
It works term by term, inside the sum over permutations in the
symmetrizer. Thus, without loss of generality, we may use the
simple clustering considered above. We note that on the clock face
formed by the labels $1$, \ldots, $k$ within each cluster, there
is always a factor connecting any two particles at the same
position, regardless of their spin. This factor vanishes when the
particles coincide. Since there are only $k$ distinct positions,
when $k+1$ particles come to the same point, the clock positions
must coincide in at least two cases, so that the wave function
vanishes, which completes the proof.

We do not have a direct general proof of the equality of these
explicit wave functions and the formal expressions Eq.\
(\ref{wf1}), but we have performed a number of consistency checks.
First, the wave functions are polynomials of the correct degree.
From Eq.\ (\ref{wf1}), we can infer what the total degree should
be. The parafermions of the correlator contribute with (see
\cite{gep1}) $-1 \cdot 2kp \cdot (1-\frac{1}{k})$. The factors of
the 2,2,1 part are $2 \cdot \frac{2}{k} \cdot \frac{1}{2}
kp(kp-1)$ and $1 \cdot \frac{1}{k} \cdot (kp)^2$. Adding these
gives, for $M=0$, $pk(3p-2)$. We need to check whether Eq.\
(\ref{wf11}) gives the same degree. For the $i$th up particle, the
degree of $z_i^\uparrow$ in the product of up-up factors is
$2(p-1)$, and in the up-down factors is $p$. Thus the net degree
in $z_i^\uparrow$ is $N_\phi=3p-2=3N/2k-2$, or for general $M$,
$N_\phi=3p+M(N-1)-2=(M+3/2k)N-2-M$. This gives the filling factor
$\nu=2k/(2kM+3)$ \cite{as1}, which reduces to that for the
Halperin states for $k=1$, and also the shift, defined as
$N_\phi=N/\nu-{\mathcal S}$, which here is ${\mathcal S}=M+2$ on
the sphere (for more on the shift, see Ref.\ \cite{we1}). Finally,
the total degree is $N/2$ times that in $z_i^\uparrow$, namely
$kp(3p-2)$ for $M=0$, the same as for the correlator.
Also, the numerical work described in Sec.\ \ref{numerics} below
confirms that the ground state of the appropriate Hamiltonian on
the sphere for $k=2$, $M=1$ at the given number of flux does have
a unique spin-zero ground state at zero energy, so that the
correlator and the wave function constructed above must coincide.
This also implies that the wave functions above must be spin
singlet.

\section{Correlators corresponding to states with quasiholes}
\label{qhcor}

To obtain wave functions for NASS states with quasiholes, one
inserts corresponding operators into the correlator that
corresponds to the ground state wave function. Here we will give
the form of the correlators, using standard CFT techniques, but
not the wave functions. The operators we insert have the form of a
spin field times a vertex operator, similar to the RR states
\cite{rr2} (note that the term ``spin field'' is traditional, and
has no relation to the SU($2$) ``spin'' symmetry). In the exponent
of the vertex operators, the fundamental bosons are multiplied by
(fundamental) {\it weights} of the Lie algebra SU($3$): $\varpi_1
= (\sqrt{2}/2,\sqrt{6}/6)$, $\varpi_2 = (0,\sqrt{6}/3)$. We take
the quasiholes from the triplet ${\bf 3}$ of SU($3$). The
corresponding operators are
\begin{eqnarray}
&& C_{\varpi_1} (w^\uparrow) = \sigma_{\varpi_1}
\exp (i \varpi_1 \cdot \vec{\varphi} /\sqrt{k})(w^\uparrow)
\nonu
&& C_{-\varpi_2} (w^\downarrow) = \sigma_{-\varpi_2}
\exp (- i \varpi_2 \cdot \vec{\varphi} /\sqrt{k})(w^\downarrow) \ .
\end{eqnarray}

In order to find the wave function for general $M$, we note the
following. The two bosons $\vec{\varphi} = (\varphi_1,\varphi_2)$
can be written in terms of a (hyper-)charge and spin boson
($\varphi_c$ and $\varphi_s$ respectively) by means of a simple
rotation: $\varphi_1=\frac{\sqrt{3}}{2} \varphi_c + \frac{1}{2}
\varphi_s$ and $\varphi_2=-\frac{1}{2} \varphi_c +
\frac{\sqrt{3}}{2} \varphi_s$. The $M$-dependence is then brought
in via a rescaling of the scale associated with the charge boson
$\varphi_c$. The particle and quasihole operators for general $M$
become in terms of these bosons
\begin{eqnarray}
B^{'}_{\alpha_1} &=& \psi_1
\exp(\frac{i}{\sqrt{2k}}(\sqrt{2kM+3}\,\varphi_c+\varphi_s))
(z^\uparrow) \label{elup} \\
B^{'}_{-\alpha_2} &=& \psi_2
\exp(\frac{i}{\sqrt{2k}}(\sqrt{2kM+3}\,\varphi_c-\varphi_s))
(z^\downarrow) \label{eldown} \\
C^{'}_{\varpi_1} &=& \sigma_\uparrow
\exp(\frac{i}{\sqrt{2k}}(\frac{1}{\sqrt{2kM+3}}\,\varphi_c+\varphi_s))
(w^\uparrow) \label{qhup} \\
C^{'}_{-\varpi_2} &=& \sigma_\downarrow
\exp(\frac{i}{\sqrt{2k}}(\frac{1}{\sqrt{2kM+3}}\,\varphi_c-\varphi_s))
(w^\downarrow) \label{qhdown} \ ,
\end{eqnarray}
where we have written $\psi_{\alpha_1}=\psi_1$,
$\psi_{-\alpha_2}=\psi_2$ , $\sigma_{\varpi_1}=\sigma_\uparrow$
and $\sigma_{-\varpi_2}=\sigma_\downarrow$ for simplicity. The
most basic spin fields $\sigma_{\uparrow,\downarrow}$ transform as
a doublet of the SU(2) subalgebra we identify with the spin of the
particles. Also, the hypercharge of the quasihole operators has
the same sign as that of the particle operators. This implies that
these are indeed quasiholes, as in earlier cases \cite{mr1,rr2};
the wave functions are nonsingular as any particle coordinate
$z_i$ approaches any quasihole coordinate $w_j$.

Note that when these operators are used in the CFT correlator
(together with a suitably chosen background charge), the extra
Laughlin factor is automatically generated. The correlator for the
component of the wave function with $N_{\uparrow,\downarrow}$
spin-up and down particles and $n_{\uparrow,\downarrow}$ spin-up
and down quasiholes is given by \widetext \top{-2,8cm}
\begin{eqnarray}
&& \wtPsi_{\rm NASS,qh}^{k,M} = \lim_{z_\infty \to \infty}
(z_\infty)^a
\langle
\exp \left( \frac{-i}{\sqrt{2k}} \left\{
[\sqrt{2kM+3}(N_\uparrow+N_\downarrow)
+\frac{n_\uparrow+n_\downarrow}{\sqrt{2kM+3}}]\varphi_c
+ [N_\uparrow-N_\downarrow+n_\uparrow-n_\downarrow]\varphi_s
\right\} \right) (z_\infty)
\nonu && \times \label{qhcrl}
C^{'}_{\varpi_1}(w_1^\uparrow) \ldots
C^{'}_{\varpi_1}(w^\uparrow_{n_\uparrow})
C^{'}_{-\varpi_2}(w_1^\downarrow) \ldots
C^{'}_{-\varpi_2}(w^\downarrow_{n_\downarrow})
B^{'}_{\alpha_1}(z_1^\uparrow) \ldots
B^{'}_{\alpha_1}(z^\uparrow_{N_\uparrow})
B^{'}_{-\alpha_2}(z_1^\downarrow) \ldots
B^{'}_{-\alpha_2}(z^\downarrow_{N_\downarrow})
\rangle \ .
\end{eqnarray}
The value of $a$ will be given momentarily. In the wave function
(\ref{qhcrl}) we inserted the most general background charge
required for neutrality in the Cartan subalgebra of SU($3$).
However, the background charge can involve only the charge boson
$\varphi_c$, which corresponds to the spin-independent background
magnetic field in the QH problem. Thus we find the condition
\begin{equation}
N_\uparrow + n_\uparrow = N_\downarrow + n_\downarrow,
\label{cond1}
\end{equation}
which is part of the requirement of SU(2) symmetry for the
correlator. The correlator is a spin-singlet, which means that the
wave function {\em for the particles} is a nonzero spin state of
the particles, with spin determined by the quasiholes.
Effectively, the quasiholes carry spin 1/2 which is added to the
spin-singlet ground state. Note that a quasihole labeled up
carries a spin down from the latter point of view, by Eq.\
(\ref{cond1}), just as it carries negative charge (hence the term
quasihole), since there is a deficiency of particles at its
location. For $N=N_\uparrow+N_\downarrow$ sufficiently large,
$N\geq n$ in fact, the spin (up or down) for each quasihole can be
chosen freely, as we will see in some examples. Using condition
(\ref{cond1}), we can calculate that $a$ must be
\begin{equation}
a=\frac{2kM+3}{2k}\left(N+\frac{n}{2kM+3}\right)^2,
\end{equation}
where $n=n_\uparrow+n_\downarrow$, in order that the limit
$z_\infty\to\infty$ exists and is nonzero.

By working out the vertex-operator part, we arrive at the
following form
\begin{eqnarray}
\lefteqn{  \wtPsi_{\rm NASS,qh}^{k,M}
(z_1^\uparrow,\ldots,z_{N_\uparrow}^\uparrow;
z_1^\downarrow,\ldots,z_{N_\downarrow}^\downarrow;
w_1^\uparrow,\ldots,w_{n_\uparrow}^\uparrow;
w_1^\downarrow,\ldots,w_{n_\downarrow}^\downarrow) = } \nonu &&
\langle \sigma_\uparrow(w_1^\uparrow) \ldots \sigma_\uparrow
(w_{n_\uparrow}^\uparrow) \sigma_\downarrow(w_1^\downarrow) \ldots
\sigma_\downarrow(w_{n_\downarrow}^\downarrow)
\psi_1(z_1^\uparrow) \ldots \psi_1(z_{N_\uparrow}^\uparrow)
\psi_2(z_1^\downarrow) \ldots \psi_2(z_{N_\downarrow}^\downarrow)
\rangle \nonu && \times \left[ \wtPsi_{\rm
H}^{(2,2,1)}(z_i^\uparrow;z_j^\downarrow) \right]^{1/k} \,
\wtPsi^M_{\rm L}(z_i^\uparrow;z_j^\downarrow) \prod_{i,j}
(z^\uparrow_i-w^\uparrow_j)^{\frac{1}{k}} \prod_{i,j}
(z^\downarrow_i-w^\downarrow_j)^{\frac{1}{k}} \nonu && \times
\label{qhnass} \prod_{i<j}
(w^\uparrow_i-w^\uparrow_j)^{\frac{1}{2kM+3} (\frac{2}{k}+M)}
\prod_{i,j} (w^\uparrow_i-w^\downarrow_j)^{\frac{-1}{2kM+3}
(\frac{1}{k}+M)} \prod_{i<j}
(w^\downarrow_i-w^\downarrow_j)^{\frac{1}{2kM+3} (\frac{2}{k}+M)}
\ .
\end{eqnarray}
\bottom{-2,7cm} \narrowtext \noindent Note that the correlator is
non-zero only if the parafermion and spin fields can be fused to
yield the identity operator.

The number of magnetic flux $N_\phi$ seen by any particle is found
to be
\begin{equation}
N_\phi=\frac{2kM+3}{2k}N+\frac{1}{2k}n -(M+2), \label{Nphi}
\end{equation}
where we used Eq.\ (\ref{cond1}). Since $N_\phi$ must be an
integer (so that the wave function is a polynomial in the
$z_i$'s), this gives another condition, that $(3N+n)/2$ [which is
an integer by Eq.\ (\ref{cond1})] must be divisible by $k$. [For
the RR states, there is an analogous condition, $2N+n$ must be
divisible by $k$. For $k$ even, this means that $n$ is even, as in
the $k=2$ case (the MR state) \cite{mr1}. In Ref.\ \cite{rr2},
only the case $n$ and $N$ both divisible by $k$ was considered.]
From Eq.\ (\ref{Nphi}), we can deduce that the quasiparticle
charge is $1/(2kM+3)$. This corresponds to a fractional flux,
$1/2k$ of the usual flux quantum. In effect, the flux quantum has
been reduced by $1/k$ by the formation of clusters, as in paired
states and in spin-polarized RR states \cite{rr2}. The factor of
$1/2$ is present already in the Halperin $k=1$ case. So if $k$ is
not divisible by $3$, the quasihole charge is $1/Q$ ($Q$ is the
denominator of the filling factor, defined in Sec.\ \ref{cft}), as
in many other cases, but if $k$ is divisible by 3, the quasihole
charge is $1/3Q$. This is similar to what happens in the MR and RR
states, where the quasihole charge is further fractionalized
(smaller than $1/Q$) when $k$ is divisible by $2$ \cite{rr2}.

The conditions (\ref{cond1}) and that $N_\phi$ be integer are
clearly necessary, but in fact are also sufficient, to ensure that
the quasihole wave functions are nonzero polynomials in the
$z_i$'s, except in the special case $n=1$ where the function
vanishes. To see this, one must examine the fusion rules for the
parafermion system, and check that the fields can be fused to the
identity operator under the stated conditions. This will be
considered in the next Section.

For completeness, we give the conformal dimensions of the particle
and quasihole operators $B^{'}_\alpha(z)$ and $C^{'}_\varpi(w)$.
To obtain these, we need the dimensions of the parafermionic and
spin fields, which are \mbox{$\Delta_\psi = 1-\frac{1}{k}$} and
\mbox{$\Delta_\sigma = \frac{k-1}{k(k+3)}$}, respectively
\cite{gep1}. Using these, we find (see also \cite{as1})
\mbox{$\Delta_{\rm part} = \frac{M+2}{2}$} and \mbox{$\Delta_{\rm
qh} = \frac{(5k-1)M+8}{2(k+3)(2kM+3)}$}.

We can show that the quasihole states we have obtained are
zero-energy eigenstates for the $k+1$-body Hamiltonian above, as
follows. We again concentrate on the case $M=0$. The argument
using the ope's of the currents $B_{\uparrow,\downarrow}(z)$ again
applies \cite{rr2}, as long as the $k+1$ $z_i$'s are brought to a
point that does not coincide with a quasihole coordinate $w_j$. To
complete the argument, we must also consider the case where the
latter condition does not hold. There are two ways to do this. One
is to note first that, as a function of the $z_i$'s for fixed
$w_j$'s, the correlator is a polynomial, as it must be to be a
valid QHE wave function. (It is {\em not} a polynomial in the
quasihole coordinates $w_j$). This is because we chose to examine
quasiholes rather than the opposite charge objects. Then the fact
that it vanishes when $k+1$ $z_i$'s coincide away from a quasihole
coordinate $w_j$ also implies that it vanishes when they are at a
$w_j$, by continuity, which holds because the function is a
polynomial in the $z_i$'s.

A second argument is also instructive. We may generalize the
argument using the current algebra null vectors to directly
address the limit of $B$'s approaching a $w_j$. There is a
generalization of the central equation for this case,
\begin{equation}
(B_{\uparrow,-1})^k|\uparrow\rangle=0. \label{null2}
\end{equation}
Here the state $|\uparrow\rangle=C_{\varpi_1}(0)|0\rangle$ is the
state in radial quantization corresponding to the quasihole
operator at $0$. There are similar equations, with successively
lower exponents, for the higher-order quasiholes (with multiples
of the charge of the basic one) obtained by successively fusing
quasiholes together. These are the null vector equations for the
highest weights in distinct representations of the affine Lie
algebra (or for the distinct primary fields) \cite{gepw,fz}. We
want to emphasize that the equation says that for certain choices
of the spins, the wave function vanishes when only $k$ particles
come to the same point (or fewer for the higher-order quasiholes).
This is a generalization of the fact that the Laughlin quasihole
is defined as the factor $\prod_i(z_i-w)$ which vanishes when any
one particle approaches $w$. (This generalization applies already
to the spinless RR states.) It is also a generalization of the
Halperin case, where a quasihole is a factor
$\prod_i(z_i^\uparrow-w^\uparrow)$ which vanishes when a single up
particle goes to $w^\uparrow$, but not when a single down particle
does. When $k$ basic quasiholes are fused at $w$ (taking the
leading term at each fusion), the null vector equations state that
the function does vanish when a single particle of appropriate
spin approaches $w$, so we have a Laughlin- (or rather Halperin-)
type quasihole in that case, as was already known for the MR
($k=2$ RR) state, for example. Note that the above spin $k/2$
state is the highest weight in a multiplet, so there is a set of
$k+1$ such null vectors in total. This does not include all
possible spin choices, as we have pointed out. To complete the
proof that the function vanishes when any $k+1$ $z_i$'s (i.e.\
$k+1$ particles of any spin) come to $w_j$, we must show that, for
all $m$,
\begin{eqnarray}
(B_{\downarrow,-1})^m(B_{\uparrow,-1})^{k+1-m}|\uparrow\rangle&=&0\nonu
(B_{\downarrow,-1})^m(B_{\uparrow,-1})^{k+1-m}|\downarrow\rangle&=&0.
\end{eqnarray}
This can be done by an elementary argument, applying the SU(2)
lowering operator to Eq.\ (\ref{null2}), then another
$B_{\uparrow,-1}$, using the same equation, and then lowering
further, and so on.

\section{Fusion rules}
\label{fusions}

From now on, we focus mainly on the case $k=2$, which is a
spin-singlet analogue of the MR state. The numerical studies
reported in Section \ref{numerics} were all performed for this
special case. Analytical results for $k\geq 2$ will be presented
elsewhere \cite{as2}.

As pointed out in the introduction, the non-trivial fusion rules
play a crucial role in the ground state degeneracies. In fact, the
correlator in Eq.\ (\ref{qhnass}) does not represent a single wave
function, because in general there is more than one way in which
the spin fields and parafermion fields can be fused to the
identity. To show how this works, we will give the fusion rules,
and explain that they can be written in terms of a Bratteli
diagram. By using the correspondence between the fields of the
parafermion theory and fields of the corresponding
Wess-Zumino-Witten models (see \cite{gep1}), one finds the fusion
rules listed in Table \ref{fusrul}.

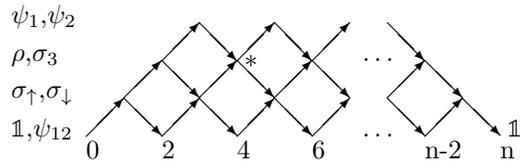
\begin{figure}[h]
\setlength{\unitlength}{1mm}
\begin{picture}(60,20)(-17,-3)
\put(0,0){\vector(1,1){5}}
\put(5,5){\vector(1,1){5}}
\put(5,5){\vector(1,-1){5}}
\put(10,0){\vector(1,1){5}}
\put(10,10){\vector(1,1){5}}
\put(10,10){\vector(1,-1){5}}
\put(15,5){\vector(1,1){5}}
\put(15,5){\vector(1,-1){5}}
\put(15,15){\vector(1,-1){5}}
\put(20,0){\vector(1,1){5}}
\put(20,10){\vector(1,1){5}}
\put(20,10){\vector(1,-1){5}}
\put(25,5){\vector(1,1){5}}
\put(25,5){\vector(1,-1){5}}
\put(25,15){\vector(1,-1){5}}
\put(30,0){\vector(1,1){5}}
\put(30,10){\vector(1,1){5}}
\put(30,10){\vector(1,-1){5}}
\put(35.5,0){\ldots}
\put(35.5,10){\ldots}
\put(40,5){\vector(1,1){5}}
\put(40,5){\vector(1,-1){5}}
\put(40,15){\vector(1,-1){5}}
\put(45,0){\vector(1,1){5}}
\put(45,10){\vector(1,-1){5}}
\put(50,5){\vector(1,-1){5}}
\put(0,-3){0}
\put(10,-3){2}
\put(20,-3){4}
\put(30,-3){6}
\put(45,-3){n-2}
\put(55,-3){n}
\put(-10,0){$\id$,$\psi_{12}$}
\put(-10,5){$\sigma_\uparrow$,$\sigma_\downarrow$}
\put(-10,10){$\rho$,$\sigma_3$}
\put(-10,15){$\psi_1$,$\psi_2$}
\put(56,0){$\id$}
\put(21,8.1){*}
\end{picture}\\
\caption{Bratteli diagram for the spin fields of
         SU$(3)_2/[$U(1)$]^2$.}
\label{brat}
\end{figure}

An examination of the fusion rules shows that there are different
cases, according to the parity, even or odd, of $n_\uparrow$,
$n_\downarrow$, $N_\uparrow$, $N_\downarrow$. In the case where
all four numbers are even, the spin fields and the parafermions
can be fused to the identity separately. In the case where all
four are odd, the spin fields and the parafermions can be fused to
$\psi_{12}$ separately, and these two $\psi_{12}$'s can then be
fused to the identity. Because the quasiholes only involve the
$\sigma_{\uparrow,\downarrow}$ fields, we in fact only need the
first two columns of Table \ref{fusrul}. With this restriction,
the fusion rules can be written in terms of a Bratteli diagram,
see Figure \ref{brat}.
Each arrow stands for either a $\sigma_\uparrow$ or
$\sigma_\downarrow$ field. The arrow starts at a certain field
which can only be one of the fields on the left of the diagram at
the same height. This last field is fused with the one
corresponding to the arrow, while the arrow points at a field
present in this fusion. As an example, the arrows starting at the
$*$ are encoding the fusion rules $\rho \times \sigma_{\uparrow
(\downarrow)} = \psi_{2 (1)} + \sigma_{\uparrow (\downarrow)}$ and
$\sigma_3 \times \sigma_{\uparrow (\downarrow)} = \psi_{1 (2)} +
\sigma_{\downarrow (\uparrow)}$. One checks that the diagram is in
accordance with the first two columns of Table \ref{fusrul}. The
symbol $\id$ at the right hand side of Figure \ref{brat} indicates
that in the end we have fused the fields to the identity. This is
possible only when $n_\uparrow$ and $n_\downarrow$ are both even;
in the case where both numbers are odd, $\psi_{12}$ is obtained at
that position in the diagram. In the remaining two cases, where
$n=n_\uparrow+n_\downarrow$ is odd, we can draw a similar diagram
with the last point at the top, representing $\psi_1$ or $\psi_2$
(except when $n=1$, to which we return in a moment). In these
cases, $N=N_\uparrow+N_\downarrow$ must also be odd, in order that
the fusion of the parafermions $\psi_1$ and $\psi_2$ for the
particles can produce the appropriate field which can fuse with
the result of the spin field fusions to finally give the identity.
In the case $n=1$, it is not possible to fuse the spin fields to
obtain a parafermion, and the correlator vanishes.

For the counting formula we need to know the number of ways in
which a given number of spin fields can be fused to give a field
in the parafermion sector, that is $1$, $\psi_1$, $\psi_2$, or
$\psi_{12}$. This number equals the number of paths (of length the
number of spin fields) on the Bratteli diagram leading to the
corresponding point on the diagram. One finds that when the
numbers of spin-up and down quasiholes are $n_\uparrow$ and
$n_\downarrow$, respectively, this number, the number of fusion
channels, is $d_{n_\uparrow,n_\downarrow}=f_{n - 2}$, where the
Fibonacci number is defined by $f_m=f_{m-1}+f_{m-2}$, with the
initial conditions $f_0=1$ and $f_1=1$. This is valid for $n\geq
2$ in all four cases of $n_\uparrow$, $n_\downarrow$ even or odd,
while for $n=1$, $d_{n_\uparrow,n_\downarrow}=0$, and for $n=0$,
$d_{0,0}=1$. The result is obtained by examining the Bratelli
diagram and seeing that the number of paths obeys the recurrence
relation that defines the Fibonacci number. We note that this
result, the Fibonacci number $f_{n-2}$, is the same as for $n$
quasiholes in the $k=3$ RR states \cite{rr2}. This is a
manifestation of level-rank duality, here between SU$(3)_2$ and
SU$(2)_3$.

The final fusion of the spin fields with the parafermions from the
particles must produce the identity, in order that the correlator
be nonzero. When $d_{n_\uparrow,n_\downarrow}$ is nonzero,
necessary and sufficient conditions for this are $N_\uparrow +
n_\uparrow \equiv N_\downarrow + n_\downarrow$ (mod 2) and
$N_\uparrow - n_\downarrow \equiv 0$ (mod 2). The first condition
is the mod 2 version of the condition (\ref{cond1}), while the
second is equivalent to the condition $(3N+n)/2\equiv 0$ (mod 2),
on using Eq.\ (\ref{cond1}). Note that, as can be inferred from
the fusion rules for the $\psi$ fields (see Table \ref{fusrul}),
the fusion of the $\psi$ fields from the particle operators does
not increase the degeneracy. The Abelian properties of the $\psi$
fields correspond to the Abelian statistics (Bose or Fermi) of the
underlying particles.

These results give the degeneracy $d_{n_\uparrow,n_\downarrow}$ of
quasihole states for fixed positions and spins of the quasiholes,
which is the basis for non-Abelian statistics properties. We see
that the result does not depend on how many quasiholes are spin up
or spin down, and the sum over all choices of spins gives a
further factor of $2^n$ spin degeneracy for sufficiently large
$N$, when only the positions are fixed. In the following Sections,
we examine the total degeneracies of quasihole states when both
their positions and spins are unrestricted. These are more
suitable for numerical checks, and are finite numbers on the
sphere (for a disk in the plane, they are infinite, and contain
information about edge excitations as well as bulk quasiholes). As
in cases studied earlier \cite{rr1,rr2}, the total degeneracies
are not just the numbers found above times a factor for the
spatial degeneracy contribution, but involve partitioning the
Fibonacci numbers above into a sum of positive integers. We also
note here that when a generic Hamiltonian has a ground state in
the NASS phase, the degeneracies will not be exact, but will be as
given in this Section when all quasiholes are asymptotically far
separated. This will not be considered further in this paper.

\section{Spatial degeneracies}
\label{sphere}

Techniques for calculating degeneracies for a spherical geometry
are described in full detail in \cite{rr1}. On the sphere, the
relation between the number of particles and the number of flux
quanta for the ground state is given by $N_\phi = \nu^{-1} N
-\mathcal{S}$. By increasing the number of flux quanta at fixed
$N$, quasiholes are created. Moreover, when flux have been added,
there may be zero-energy states with $N$ not satisfying the
conditions required in the ground state, for example that $N$ be
even in the MR state \cite{rr1}. In general, we would define the
number of flux ``added'' as $\Delta N_\phi=N_\phi-\nu^{-1}N+{\cal
S}$. This is defined relative to a ground state at the same number
of particles, even though such a zero-energy state for $N$ not
divisible by $2k$ (or $k$ for the RR states) would require a
non-integer number of flux, and does not exist. Consequently,
while our $N_\phi$ is always an integer, as discussed in Sec.\
\ref{qhcor}, $\Delta N_\phi$ does not have to be an integer. The
number of quasiholes $n$ can then be defined as $n=k\Delta N_\phi$
(for the RR states), or $n=2k \Delta N_\phi$ for the NASS states
considered here, in agreement with our formula for $N_\phi$ in
Eq.\ (\ref{Nphi}) above.

To explain the spatial degeneracies, we use the Laughlin case as
an example, and follow the discussion of \cite{rr1}. The Laughlin
wave function for $N$ particles in the presence of $n$ quasiholes
can be written as \cite{laugh}
\begin{eqnarray}
&& \wtPsi_{\rm L,qh}^M(z_1,\ldots,z_N;w_1,\ldots,w_n) = \nonu
&& \prod_{i<j}(z_i-z_j)^M \prod_{i,k} (z_i-w_k) \ .
\end{eqnarray}
For this state, $n = \Delta N_\phi$ (adding one quantum of flux
creates one quasihole), so we have
$N_\phi = M(N-1)+n$, where we used that ${\cal S}=M$
for the Laughlin state.
To continue, we calculate the degeneracy due to the
presence of the quasiholes, by expanding
the factor $\prod_{i,k} (z_i-w_k)$ in sums of products
of the elementary symmetric polynomials
\begin{equation}
e_m = \sum_{i_1<i_2<\dots<i_m} w_{i_1} w_{i_2}\dots w_{i_m} \ .
\end{equation}
Viewing the coordinates $w_i$ as coordinates of bosons, we find
that $n$ bosons are to be placed in $N+1$ orbitals. The dimension
of the space of available states (linearly-independent wave
functions) equals the number of ways in which one can put $n$
bosons in $N+1$ orbitals, which is
\begin{equation}
\left( \begin{array}{c}
N + n \\ n \end{array} \right) \ .
\end{equation}
This is the spatial degeneracy we are after, although for the
simple case of the Laughlin states.

The situation for the MR state is discussed in detail in
\cite{rr1}. For the MR state, there is an additional complication,
namely quasihole states in which there are unpaired fermions are
possible; this is the origin of the degeneracies for fixed
quasihole positions, already discussed. We will denote the number
of unpaired particles by $F$, with the requirement that $N-F$ be
even, so that the number of unbroken pairs $(N-F)/2$ is an
integer. For $n$ sufficiently large, $N$ need not be even in the
zero-energy states. It was found that the spatial degeneracy
depends on the number of unbroken pairsx; in fact, for the MR
state, it was given by \cite{rr1}
\begin{equation}
\left( \begin{array}{c}
\frac{N-F}{2} + n \\ n \end{array} \right) \ .
\end{equation}

For the clustered state of \cite{rr2} (in which the particles form
clusters of order $k$ rather than pairs), the spatial degeneracy
is given similarly by \cite{rr2,gr1}
\begin{equation}
\left( \begin{array}{c}
\frac{N-F}{k} + n \\ n \end{array} \right) \ ,
\end{equation}
where $N-F$ must be divisible by $k$ (there appears to be no known
general analytic proof of this formula for $k>2$).

Based on these earlier results, and because the NASS wave
functions involve clusters of up particles and separately of
downs, we expect that the spatial degeneracy for the NASS states
is just a product of two binomial coefficients, involving the
$F_1$, $F_2$ ``unclustered'' particles (or parafermions) for spins
up and down, respectively:
\begin{equation} \label{ordeg}
\left( \begin{array}{c} \frac{N_\uparrow-F_1}{k} + n_\uparrow \\
n_\uparrow
\end{array} \right)
\left( \begin{array}{c} \frac{N_\downarrow-F_2}{k} + n_\downarrow
\\ n_\downarrow
\end{array} \right) \ .
\end{equation}
Again, $N_\uparrow-F_1$ and $N_\downarrow-F_2$ must be divisible
by $k$. Notice that these numbers never depend on $M$.

The case $k=1$, $F_1=F_2=0$ gives the spatial degeneracies for the
general Halperin $(m',m',m)$ Abelian states (as was mentioned
briefly in Ref.\ \cite{rr1} for a particular case). For the
spin-singlet cases $(m+1,m+1,m)$ ($m=M+1$) of interest here, the
conditions that correspond to zero-energy states with only the
particle and quasihole numbers, $N$ and $n$, fixed are that
$N_\uparrow+n_\uparrow=N_\downarrow+n_\downarrow$. For $N$
sufficiently large, these allow any choice of spin, up or down,
for each quasihole, giving a factor of $2^n$ spin degeneracy for
fixed positions (this holds for all $k$). Such degeneracy does not
contribute to the degeneracy on which non-Abelian statistics
depends, and the statistics is Abelian in the present $k=1$ case
(as of course was expected). The full degeneracy in this case
$k=1$ is obtained by summing (\ref{ordeg}) over $N_\uparrow$,
$N_\downarrow$, $n_\uparrow$, $n_\downarrow$, subject to the
conditions just stated, with $F_1=F_2=0$. These imply that the
summation is over only the possible values of
$S_z=(N_\uparrow-N_\downarrow)/2$, and the result is
\begin{equation}
\left(\begin{array}{c}N+n\\n\end{array}\right).
\end{equation}
Note that this number includes the spin degeneracy. If we take the
ratio to the number
\begin{equation}
\left(\begin{array}{c}(N+n)/2\\n\end{array}\right)
\end{equation}
of quasihole states with, say, $n_\downarrow=0$, the result tends
to $2^n$ as $N\to\infty$, which is again the spin degeneracy for
fixed positions. In the remainder of this paper, we concentrate
exclusively on $k=2$.

With the orbital degeneracies in hand, we need to know how to
break up the degeneracies stemming from the fusion rules of the
spin fields. So in fact we need to know the number of unpaired
particles of either spin, $F_1$, $F_2$, for each possible path on
the Bratteli diagram. The next Section will treat this problem, by
partitioning the numbers $d_{n_\uparrow,n_\downarrow}$ in the
following way
\begin{equation}
d_{n_\uparrow,n_\downarrow} =
f_{n_\uparrow+n_\downarrow-2}=\sum_{F_1,F_2}\left\{
\begin{array}{cc} n_\uparrow & n_\downarrow \\
F_1 & F_2 \\ \end{array} \right\}_2 \ .
\end{equation}
The symbol $\{^{n_\uparrow n_\downarrow}_{F_1 F_2}\}_2$ is
interpreted as the number of zero-energy states containing
$n_\uparrow$, $n_\downarrow$ quasiholes at fixed positions and
$F_1$, $F_2$ unpaired parafermions. The symbol vanishes if the
conditions that $F_1-n_\downarrow$, $F_2-n_\uparrow$
be even are not satisfied; these conditions are equivalent to the
two mod 2 conditions discussed at the end of Sec.\ \ref{fusions}.
The subscript $2$ indicates that we are dealing with the case
$k=2$.

\section{Counting SU$(3)_2/[$U$(1)]^2$ parafermion states}
\label{parafermions}

It was explained in \cite{rr1} that the state counting for the MR
state involves the systematics of Majorana fermions, which act as
BCS quasiparticles (unpaired fermions) occupying zero-energy
states \cite{rg}. For the more general RR (spin-polarized) states
with order $k$ clustering \cite{rr2}, the Majorana fermion is
replaced by an SU($2)_k/$U($1)$ parafermion \cite{rr2,gr1}.

We recall that for the NASS state at level $k=2$, the role of the
Majorana fermion is taken over by the parafermions that are
associated to SU$(3)_2/[$U(1)$]^2$. The (spin-up or down)
quasiholes correspond to (two different) spin fields
$\sigma_{\uparrow,\downarrow}$ of this parafermion theory (see
Section \ref{qhcor}). In Section \ref{fusions}, we calculated the
number of different quantum states (conformal blocks for the
correlators) that can result from introducing $n_\uparrow$
$\sigma_\uparrow$ and $n_\downarrow$ $\sigma_\downarrow$ spin
fields (and also varying the number of particles $N_\uparrow$,
$N_\downarrow$) with the result
$d_{n_\uparrow,n_\downarrow}=f_{n_\uparrow+n_\downarrow-2}$. The
degeneracy results from the presence of varying numbers of
particles that are not members of clusters, which in the
correlators can be identified with the fundamental parafermions
$\psi_1$, $\psi_2$ of the parafermion theory. $F_1$ ($F_2$) is the
number of $\psi_1$ ($\psi_2$) excitations. These numbers are
subject to the conditions that $F_1\equiv n_\downarrow$ (mod 2),
$F_2 \equiv n_\uparrow$ (mod 2), otherwise the number of
zero-energy states is zero; these conditions come from those
discussed in Sec.\ \ref{fusions}. In the previous Section, we
found that the orbital degeneracy depends on the numbers $F_1$ and
$F_2$, see Eq.\ (\ref{ordeg}). We now turn to the calculation of
the symbols
\begin{equation}
\left\{ \begin{array}{cc}
n_\uparrow & n_\downarrow \\
F_1 & F_2 \\
\end{array} \right\}_2 \ ,\label{symbol2}
\end{equation}
which partition the degeneracies $d_{n_\uparrow,n_\downarrow}$ in
the correct way. We will also keep track of the angular momentum
($L$) multiplet structure associated with these parafermion
states. To do this, we have to go through a series of steps.

First of all, we consider the infinite (chiral) character
corresponding to the full parafermionic CFT
(see \cite{pfchar})
\begin{equation}
{\rm ch}(x_1,x_2;q) = \sum_{F_1,F_2}
\frac{q^{(F_1^2+F_2^2-F_1F_2)/2}} {(q)_{F_1}(q)_{F_2}} x_1^{F_1}
x_2^{F_2} \ ,
\end{equation}
where $(q)_a= \prod_{j=1}^a (1-q^j)$ for integer $a$. Here $F_1$
and $F_2$ are unrestricted non-negative integers, and $x_1$,
$x_2$, $q$ are indeterminates.

What is needed for our purposes here is the truncation of this
expression to (a sum of) polynomials
$Y_{n_\uparrow,n_\downarrow}(x_1,x_2;q)$ that describes the states
that occur when $n_\uparrow$, $n_\downarrow$ spin fields
(quasiholes) are present. The approach is described in Refs.\
\cite{sc1,bs1}, see also Ref.\ \cite{gr1}, and details for the
present case will be given in Ref.\ \cite{as2}. We find that these
polynomials satisfy the following recursion relations
\begin{eqnarray}
\nonumber
Y_{(n_\uparrow,n_\downarrow)} & = & Y_{(n_\uparrow-2,n_\downarrow)} +
x_1 q^{\frac{n_\uparrow-1}{2}}
Y_{(n_\uparrow-2,n_\downarrow+1)} \ , \\
\label{recursion}
Y_{(n_\uparrow,n_\downarrow)} & = & Y_{(n_\uparrow,n_\downarrow-2)} +
x_2 q^{\frac{n_\downarrow-1}{2}} Y_{(n_\uparrow+1,n_\downarrow-2)}
\end{eqnarray}
with initial conditions
\begin{eqnarray}
&& Y_{(1,0)}=Y_{(0,1)}=0 \ ,
\nonumber\\
&& Y_{(0,0)}=Y_{(2,0)}=Y_{(0,2)}=1 \ ,
\nonumber\\
&& Y_{(1,1)} = q^{\frac{1}{2}} x_1 x_2 \  .
\end{eqnarray}
Recursion relations similar to the above (but lacking the
$x_{1,2}$ dependence), have been considered in the mathematical
literature on special polynomials associated to SU$(3)_2$, see for
instance \cite{sw}. The coefficient of $x_1^{F_1}x_2^{F_2}$ in the
polynomial $Y_{(n_\uparrow,n_\downarrow)}$ is a polynomial in $q$
with the sum of the coefficients equal to the symbol
(\ref{symbol2}), that is
\begin{equation}
Y_{(n_\uparrow,n_\downarrow)}(x_1,x_2,1)=\sum_{F_1,F_2}x_1^{F_1}x_2^{F_2}
\left\{
\begin{array}{cc}
n_\uparrow & n_\downarrow \\ F_1 & F_2
\end{array}
\right \}_2\ .
\end{equation}
We notice that the recursion relations preserve the conditions on
the parities of $F_1$, $F_2$ (the exponents of $x_1$, $x_2$) that
are part of the definition of the symbol (\ref{symbol2}). In the
limit where $(n_\uparrow,n_\downarrow) \rightarrow
(\infty,\infty)$,  the sum of these polynomials over the four
choices, $n_\uparrow$ and $n_\downarrow$ each either even or odd,
approaches the expression ${\rm ch}(x_1,x_2;q)$ given above.

The coefficient of $x_1^{F_1}x_2^{F_2}$ in the polynomial
$Y_{n_\uparrow,n_\downarrow}$ has a special form, which allows us
to extract information on the $L$ quantum numbers of the
parafermion states: after multiplying with a factor
$q^{-(n_\uparrow F_1 + n_\downarrow F_2)/4}$, we recognize a sum
of terms of the form $q^{l_z}$, which together form a collection
of angular momentum ($L$) multiplets with quantum numbers
$L_z=l_z$ \cite{gr1}.

To illustrate the above, we present the polynomials
for the case of two added flux quanta, giving eight quasiholes.
The polynomials are
\widetext
\top{-2,8cm}
\begin{eqnarray}
\nonumber
Y_{(8,0)} & = & 1+(q^2+q^3+2q^4+q^5+q^6)x_1^2+ q^8 x_1^4
+ (q^6+q^7+q^8+q^9+q^{10})x_1^4 x_2^2 \ ,
\nonumber\\
Y_{(7,1)} & = & (q^{\frac{1}{2}} + q^{\frac{3}{2}} +
q^{\frac{5}{2}} +q^{\frac{7}{2}}) x_1 x_2
+ (q^{\frac{7}{2}} + 2 q^{\frac{9}{2}} + 2 q^{\frac{11}{2}} +
2 q^{\frac{13}{2}} + q^{\frac{15}{2}}) x_1^3 x_2
+ q^{\frac{19}{2}} x_1^5 x_2^3 \ ,
\nonumber\\
Y_{(6,2)} & = & 1 + (q^2 + q^3 + q^4) x_1^2 +
(q^2 + q^3 + 2 q^4 + q^5 + q^6) x_1^2 x_2^2
+ (q^6 + q^7 + q^8) x_1^4 x_2^2 \ ,
\nonumber\\
Y_{(5,3)} & = & (q^{\frac{1}{2}} + 2 q^{\frac{3}{2}} +
2 q^{\frac{5}{2}} + q^{\frac{7}{2}}) x_1 x_2 +
(q^{\frac{7}{2}} + q^{\frac{9}{2}} + q^{\frac{11}{2}}) x_1^3 x_2
+ (q^{\frac{9}{2}} + q^{\frac{11}{2}} + q^{\frac{13}{2}}
+ q^{\frac{15}{2}}) x_1^3 x_2^3 \ ,
\nonumber\\
Y_{(4,4)} & = & 1 + q^2 x_1^2 + q^2 x_2^2 +
   (q^2 + 2q^3 + 3q^4 + 2q^5 + q^6) x_1^2 x_2^2 + q^8 x_1^4 x_2^4 \ ,
\nonumber\\
{\rm etc.} &&
\end{eqnarray}
\bottom{-2,7cm} \narrowtext \noindent From the polynomial
$Y_{(5,3)}$ (as an example), we read off the following nonzero
values of the symbols
\begin{eqnarray}
&& \left \{
\begin{array}{cc}
5 & 3 \\
1 & 1 \\
\end{array} \right \}_2
= 6  \quad
(L=\frac{3}{2} \, , \, L=\frac{1}{2} )  \, ,
\nonumber\\
&& \left \{
\begin{array}{cc}
5 & 3 \\
3 & 1 \\
\end{array} \right \}_2
= 3
\quad  (L=1 )  \, ,
\nonumber\\
&& \left \{
\begin{array}{cc}
5 & 3 \\
3 & 3 \\
\end{array} \right \}_2
= 4
\quad (L=\frac{3}{2})  \ .
\end{eqnarray}

In fact, it is possible to write the polynomials
$Y_{(n_\uparrow,n_\downarrow)}$ in a closed form \cite{babs},
\begin{eqnarray}
Y_{(n_\uparrow,n_\downarrow)}(x_1,x_2;q) = {\sum_{F_1,F_2}}'
q^{(F_1^2+F_2^2-F_1 F_2)/2} x_1^{F_1} x_2^{F_2} \nonumber\\ \times
\left[
\begin{array}{c}
\frac{n_\uparrow+F_2}{2} \\
F_1 \\
\end{array}
\right]
\left[
\begin{array}{c}
\frac{n_\downarrow+F_1}{2} \\
F_2 \\
\end{array}
\right] \ ,
\end{eqnarray}
where $\left[ \begin{array}{c} a \\ b \end{array}\right]$ is the
$q$-deformed binomial ($q$-binomial), defined as $\left[
\begin{array}{c} a \\ b \end{array}\right] =
\frac{(q)_a}{(q)_b(q)_{a-b}}$, and the prime on the sum denotes
the restriction on $F_1$, $F_2$ values. Using the property that in
the limit $q \to 1$ the $q$-binomials become ordinary binomials,
we find the following explicit formula for $\{ \}_2$ (under the
same conditions on $F_1$, $F_2$, otherwise it vanishes):
\begin{equation} \label{symbol}
\left \{
\begin{array}{cc}
n_{\uparrow} & n_{\downarrow} \\
F_{1} & F_{2} \\
\end{array} \right \}_2 =
\left( \begin{array}{c}
\frac{n_\uparrow+F_2}{2} \\ F_1 \\
\end{array} \right)
\left( \begin{array}{c}
\frac{n_\downarrow+F_1}{2} \\ F_2 \\
\end{array} \right) \ .
\end{equation}
Note that if we take the sum over all $F_1$ and $F_2$,
we indeed find the correct value, namely a Fibonacci number
\begin{equation}
{\sum_{F_1,F_2}}' \left( \begin{array}{c} \frac{n_\uparrow+F_2}{2}
\\ F_1 \\
\end{array} \right)
\left( \begin{array}{c}
\frac{n_\downarrow+F_1}{2} \\ F_2 \\
\end{array} \right) = f_{n_\uparrow+n_\downarrow-2} \ .
\end{equation}

While Eq.\ (\ref{symbol}) gives us just the number, we can also
obtain the angular momentum content easily. The binomial $\left(
\begin{array}{c} a \\ f \\ \end{array} \right)$ is interpreted as
the number of possible ways of putting $f$ fermions in $a$ boxes
which have quantum  numbers $L_z=-(a-1)/2,\ldots,(a-1)/2$ assigned
to them. In this way, an angular momentum multiple structure is
assigned to the binomials (see \cite{rr1}). The angular momentum
content of the symbols $\{ \}_2$ is obtained by adding the angular
momenta associated to the binomials in the usual way.

\section{Final counting formula}
\label{cfor}

We are now in the position to write down the formula for the total
degeneracy of zero-energy quasihole states of the  $k=2$
non-Abelian spin-singlet states. Recall that there are two
conditions on the numbers of quasiholes (see Section \ref{qhcor}).
The first condition is $N_\uparrow + n_\uparrow = N_\downarrow +
n_\downarrow$, which implies that the correlator is a
spin-singlet, or that the wave functions have total spin
determined by the spin-1/2 quasiholes added. The other condition
was that $(3N+n)/2$ be even, to ensure that $N_\phi$ is an
integer, where $N=N_\uparrow+N_\downarrow$, and
$n=n_\uparrow+n_\downarrow = 4 \Delta N_\phi$, which relates the
number of excess flux quanta and the number of quasiholes added.
These imply that $N_\uparrow - n_\downarrow=N_\downarrow -
n_\uparrow$ must be even.

The result of the previous few Sections is now that the total
number, summed over all spin components, of zero-energy states as
a function of the number of particles and added flux quanta is
\begin{eqnarray}
&& \# (N,\Delta N_\phi) = \sideset{}{'}\sum_{N_{\uparrow,
\downarrow};n_{\uparrow,\downarrow};F_{1,2}}
\left( \begin{array}{c}
\frac{n_\uparrow+F_2}{2} \\ F_1 \\
\end{array} \right)
\left( \begin{array}{c}
\frac{n_\downarrow+F_1}{2} \\ F_2 \\
\end{array} \right)
\nonumber \\
&& \times \label{genfor}
\left(
\begin{array}{c}
\frac{N_{\uparrow}-F_{1}}{2} + n_{\uparrow}  \\
n_{\uparrow} \\
\end{array} \right)
\left(
\begin{array}{c}
\frac{N_{\downarrow}-F_{2}}{2} + n_{\downarrow}  \\
n_{\downarrow} \\
\end{array}
\right) \ ,
\end{eqnarray}
where the prime on the sum indicates that it is restricted to
values obeying all the conditions just mentioned, and to
$N_\uparrow-F_1$ and $N_\downarrow-F_2$ even as discussed in Sec.\
\ref{sphere}. Note again that these conditions imply that
$n_\uparrow+F_2$ and $n_\downarrow+F_1$ are even.

In addition, the orbital angular momentum decomposition of the
states can be obtained, by combining the angular momenta found in
the orbital and parafermion factors in the preceding two Sections.
The spin quantum number of any given state is simply
$S_z=(N_\uparrow-N_\downarrow)/2$ and one readily recognizes the
multiplet structure for the SU($2$) spin symmetry. (We remark that
the parafermion theory by itself does not have a proper SU($2$)
spin symmetry.)

In Table \ref{count1}, we present counting results for $N=4$, $8$,
$12$ and $\Delta N_\phi = 1$, $2$, $3$, $4$. We specify the number
of states as a function of the $L$ and $S$ quantum numbers. In
Table \ref{count2} we give some results for $N$ not a multiple of
four. Notice that for $n=1$, there are no zero-energy states, as
expected from Sec.\ \ref{fusions}. The results listed in Table
\ref{count1} and \ref{count2} are for the cases we checked
numerically, as we discuss in the next Section, and are in full
agreement with those results.

\section{Numerical methods and results}
\label{numerics}

We next turn to some numerical studies of the NASS states. We
consider only cases where the particles are fermions, to represent
electrons. We have studied the $k=2$, $M=1$ ($\nu=4/7$) case in
both the toroidal (PBC) and spherical geometries.  We first
present the results for the sphere. As discussed before the
flux-charge relation for this state is $N_\phi=7N/4-3$.  The
number of single-particle orbitals (the lowest LL degeneracy) is
$N_\phi+1$. In order to make contact with the results on more
conventional geometries the radius $R$ of the sphere has to be
chosen so that the number of flux is $N_\phi=2 R^2$ (where the
magnetic field strength $B$ is fixed, such that the magnetic
length is 1 in our units), so $R=\sqrt{N_\phi/2}$ \cite{hald}. The
filling factor is $\nu=N/N_\phi=2\pi \bar{n}$, where
$\bar{n}=N/(4\pi R^2)$ is the particle number density.

For numerical purposes, it is best to re-express the interaction
Hamiltonian in terms of projection operators onto different values
of the total angular momentum for different groups of particles
\cite{hald}. For the $M=1$, $k=2$ case of the NASS states, the
required Hamiltonian can be written as
\begin{equation}
H=U\sum_{i<j<k} P_{ijk}(3N_\phi/2-3,3/2) +
V'\sum_{i<j}P_{ij}(N_\phi,0),
\end{equation}
with $U$, $V'>0$. Here $P_{ijk}(L,S)$ ($P_{ij}(L,S)$) are
projection operators for the three (resp., two) particles
specified onto the given values of total angular momentum $L$ and
spin $S$ for the three (resp., two) particles. Each projection is
normalized to $P^2=P$. To see that this is the required
Hamiltonian, that corresponds to the short range $\delta$-function
interaction for $M=0$, and gives the same numbers of zero-energy
states found above, note the following. First, the maximal angular
momentum for several particles corresponds to the closest approach
of those particles \cite{hald}. In particular, the two-body term
is a contact interaction, and $V'=V_0$, the zeroth Haldane
pseudo-potential \cite{hald}. The two-body term implies that any
zero-energy states must have no component with total angular
momentum $N_\phi$ and total spin zero, which, since we are dealing
with spin 1/2 fermions, means the wave function must vanish when
any two particles coincide. The wave function must therefore
contain a factor $\wtPsi_{\rm L}^1$; multiplication by this factor
defines a one-one mapping of the full space of states of spin 1/2
bosons in the lowest LL, with $N_\phi$ reduced by $N-1$, onto the
subspace of states of the fermions that is annihilated by the
two-body term in $H$. Under this mapping, the three-body
Hamiltonian for the $M=0$ case corresponds to the three-body term
in $H$, and selects the corresponding states as zero-energy
states. In particular, the total spin of the three bosons when
they coincide (and hence of the fermions) must be $3/2$. Hence the
zero-energy eigenstates of the present Hamiltonian are given by
the results derived earlier. Note also that $H$ can be rewritten
in terms of $\delta$-functions and their derivatives. The
zero-energy eigenstates of this Hamiltonian were found for various
$N$ and $N_\phi$ values, and analyzed in terms of $L$ and $S$. The
results are shown in Tables \ref{count1} and \ref{count2}, and
agree with the counting formulas presented above.

\widetext
\top{-2,8cm}

\begin{figure}
\inseps{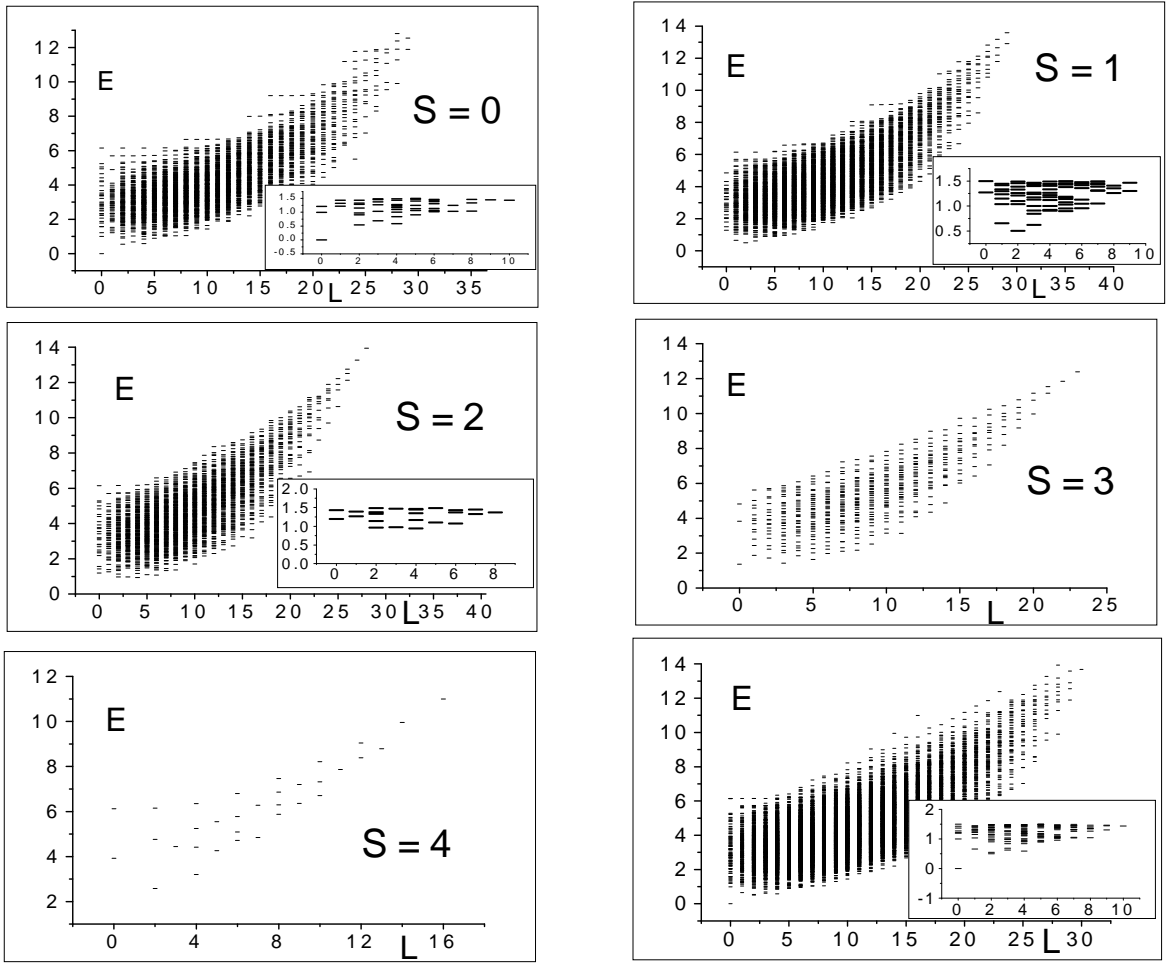}{0.646} \caption{The spectrum of the NASS model
ground state for $N=8$ and 4/7 filling. The last panel shows all
$S$ values combined. The insets are the low lying levels.}
\label{spectrum}
\end{figure}

\bottom{-2,7cm}
\narrowtext

Next we discuss the full spectra of the Hamiltonian. In Fig.\
\ref{spectrum} we show the excitation spectrum of an $N=8$,
$N_\phi=11$ system classified by the total spin $S=0$ to $S=4$.
Whenever necessary we have shown the low-lying spectrum in an
inset. The frame in the lower right hand corner shows the entire
spectrum irrespective of the total spin quantum number of the
state. The choice of $U$ and $V'$ is immaterial to the ground
state, which is always the unique zero-energy eigenstate of $H$.
Obviously, the excitation spectrum will depend on the choice of
these coefficients.  In producing Fig.\ \ref{spectrum} we chose
$U=V'=1$ so that $H$ is a sum of projection operators. There
appears to be a well-defined gap, suggesting that the system is
incompressible (for spin as well as charge) in the thermodynamic
limit, as assumed in the preceding analysis. In this connection,
we may point out that, as well as the quantized Hall conductivity
for charge, our system then has a quantized Hall conductivity for
spin, given by $k/4\pi$ in natural ($\hbar=1$) units, which is
associated with the SU$(2)_k$ subalgebra of the chiral algebra
(see e.g.\ Ref.\ \cite{rg} and references therein). Collective
modes with $S=0\ (L=2,3,4)$ and $S=1 \ (L=1,2,3)$ below the
continuum can be tentatively identified in the spectra (see insets
in Fig.\ \ref{spectrum}). That is, these may be finite-size
dispersion curves of single neutral excitations in plane-wave
(spherical harmonics on the sphere) wave functions, which would be
charge and spin modes, respectively. We shall not address the
precise nature of these neutral modes here.

\begin{figure}
\inseps{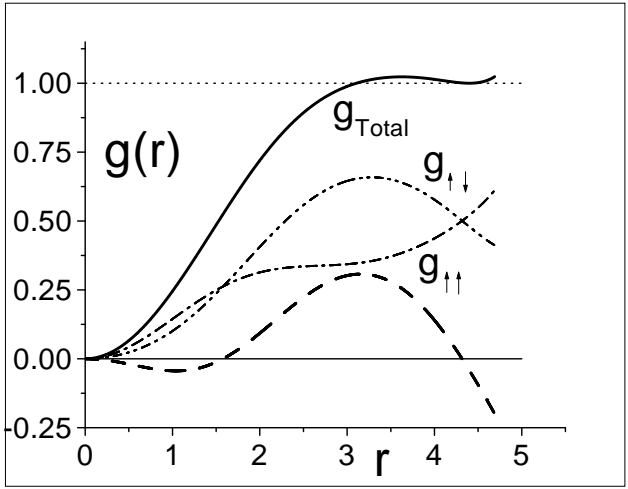}{0.346} \caption{The spin up-up and spin up-down
pair correlation functions, together with their  sum (solid line)
and difference (dashed line), versus the chord distance,
calculated in the ground state for $N=8$, $N_\phi=11$
($\nu=4/7$).}
\label{pairc}\end{figure}

In Fig.\ \ref{pairc} we show the various pair correlation
functions of interest, $g_{\uparrow \uparrow}(r)$, $g_{\uparrow
\downarrow}(r)$, as well as $g_{\rm Total}=g_{\uparrow
\uparrow}(r)+g_{\uparrow \downarrow}(r)$ and $g_{\uparrow
\downarrow}(r)-g_{\uparrow \uparrow}(r)$. The widely-different
correlations between like and opposite spins is no doubt magnified
by finite-size effects.

The 8-electron system is the first non-trivial size and is
probably too small for any meaningful comparisons of the overlap
with the 2-body Coulomb potential problem.  Nonetheless we found a
nontrivial overlap-squared (about 55\%) with the ground state of
the Coulomb potential for particles in the lowest LL (again, with
no Zeeman term), at the same $N$, $N_\phi$. By modifying the value
of the lowest pseudo-potential for the Coulomb interaction this
overlap-squared can be improved to 93\% (and probably beyond)
without any intervening phase transition (an energy gap with the
ground state is maintained at all times while the
pseudo-potentials are varied). However, in the lowest LL we do not
expect to produce a better trial wave function than one
constructed by the composite fermion (CF) method \cite{jain}, in
which two flux quanta per particle are attached (in the opposite
direction to the background magnetic field) and the resulting CFs
fill completely the first two LLs of CFs (with both spins). By
construction this is a uniform ($L=0$) spin-singlet state. We have
not constructed this state, as it occurs at a different flux for a
given $N$ ($N_\phi=7N/4$), making a direct comparison with our
NASS state difficult. We note, however, that for $N=8$ the CF
state corresponds to a spin-singlet Fermi-liquid-like state, as at
$\nu=1/2$. That is because the net effective field of the CFs is
zero for this size (states that lie in sequences for different
filling factors can coincide at finite size on the sphere, because
the shifts $\cal S$ may be different---see e.g.\ \cite{rr3}). We
expect that, as usual, this CF state will have a very large
overlap with the exact ground state of the Coulomb potential.
However, our numerical data for $N=8$ shows a much stronger cusp
at the $N_\phi$ of the NASS state than at the $N_\phi$ of the
state obtained by the hierarchy/CF construction, where in fact no
cusp can be discerned. See Fig.\ \ref{evflux} below.

\begin{figure}
\inseps{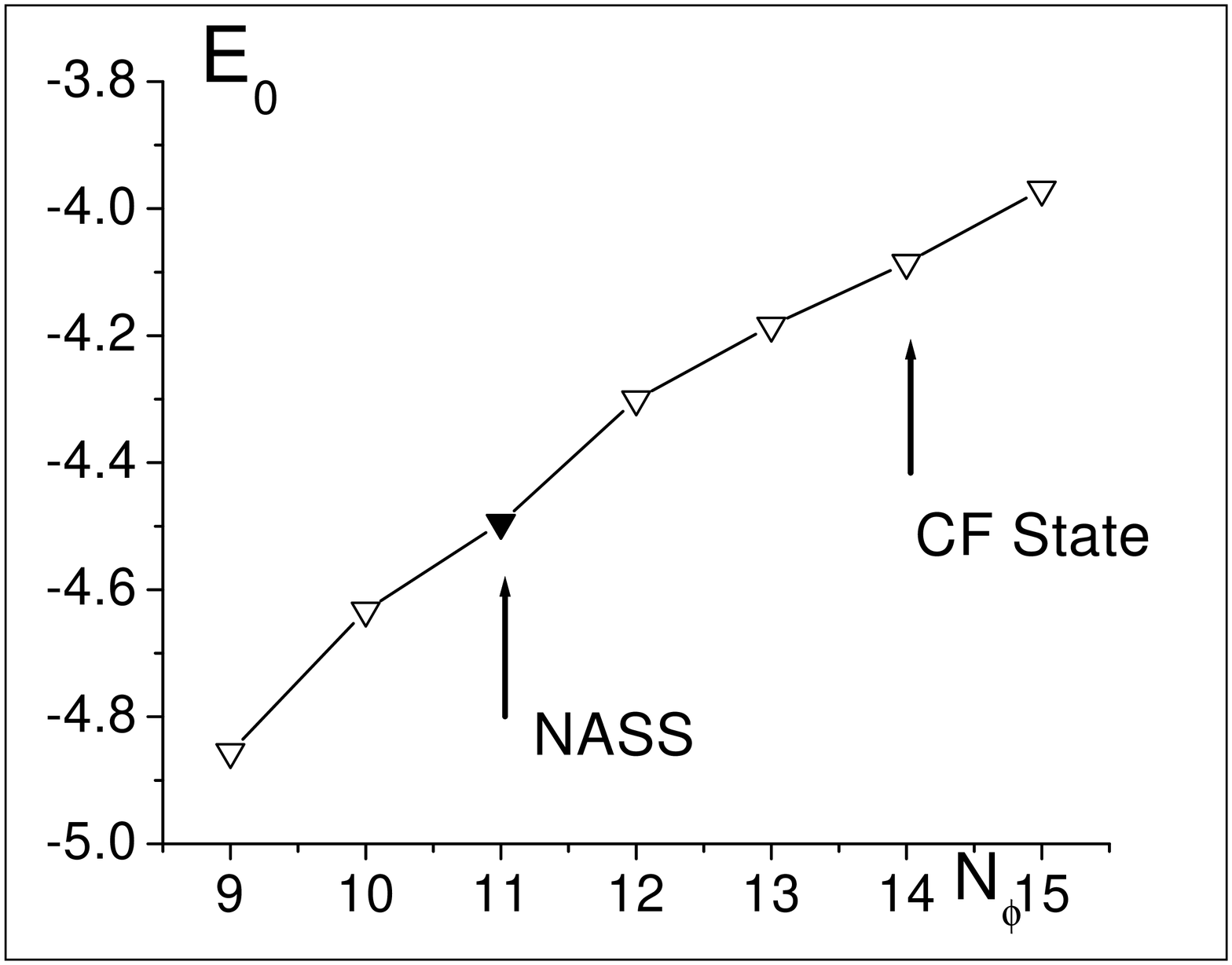}{0.346} \caption{The ground state energy of
the pure Coulomb potential in the lowest Landau level versus
$N_\phi$, at $N=8$. The numbers of flux $N_\phi$ for the NASS and
the spin-singlet CF states are marked.} \label{evflux}
\end{figure}

We have also studied the $N=8$ size on the torus.  Unfortunately,
as on the sphere this size is too small for any meaningful
comparison (e.g.\ there are only four distinct many-body $\bf k$
vectors for this size; one is at the zone center, the other three
at the zone boundary).  We would just like to point out that, for
the analog of $H$ on the torus, the degeneracy for the 4/7 state
is 2 (excluding the 7-fold center of mass degeneracy), in
agreement with the count in Sec.\ \ref{cft}, since the number of
particles is divisible by $4$. These are two ${\bf k}={\bf 0}$
states. For the pure Coulomb potential in the lowest LL the state
at $4/7$ is in fact compressible: The total spin is $S=1$ and its
$\bf k$ vector varies with the geometry of the PBC unit cell.
However, one obtains an incompressible state by increasing $V_0$
or $V_1$ and we obtained overlap-squared as large as 50\% when we
compared the lowest two states (which happen to be both $S=0$,
${\bf k}={\bf 0}$ states) to the model NASS states. Note that the
shift is zero on the torus, and there can be no interference from
$\nu=1/2$ here. We have not performed any further or systematic
studies of such issues because, as in the case of the sphere, we
suspect that the CF-based state will be closer to that of the
Coulomb potential. That is, we expect that the system with Coulomb
interaction in the lowest LL at $\nu=4/7$ is in fact in the
hierarchy/composite-fermion phase (whether spin-singlet or not),
not the NASS phase considered in this paper. We will return to
more comprehensive studies of larger sizes in the future.

\vskip 5mm

{\it Acknowledgements.} We thank P.~Bouwknegt, J.~Schliemann and
S.O.~Warnaar for discussions. The research of EA and KS is
supported in part by the Netherlands Organisation for Scientific
Research (NWO) and by the foundation FOM of the Netherlands. NR
was supported by NSF Grant No. DMR-98-18259. ER was supported by
NSF Grant No. DMR-0086191.

\vfill
\pagebreak[4]

\begin{table}
\begin{tabular}{c|c|c|c|c|c|c|c}
$\times$ & $\sigma_\uparrow$ & $\sigma_\downarrow$ &
$\sigma_3$ & $\rho$ & $\psi_1$ & $\psi_2$ & $\psi_{12}$
\\ \hline
$\sigma_\uparrow$ & $\id + \rho$ &&&&&&  \\

$\sigma_\downarrow$ & $\psi_{12}+\sigma_3$ & $\id+\rho$ &&&&&\\

$\sigma_3$ & $\psi_1 + \sigma_\downarrow$ &
$\psi_2 + \sigma_\uparrow$ & $\id + \rho$ &&&&\\

$\rho$ & $\psi_2$ + $\sigma_\uparrow$ &
$\psi_1 + \sigma_\downarrow$ & $\psi_{12} + \sigma_3$ &
$\id + \rho$ &&&\\

$\psi_1$ & $\sigma_3$ & $\rho$ & $\sigma_\uparrow$ &
$\sigma_\downarrow$ & $\id$ &&\\

$\psi_2$ & $\rho$ & $\sigma_3$ & $\sigma_\downarrow$ &
$\sigma_\uparrow$ & $\psi_{12}$ & $\id$ &\\

$\psi_{12} $ & $ \sigma_\downarrow$ & $\sigma_\uparrow$ &
$\rho$ & $\sigma_3$ & $\psi_2$ & $\psi_1$ & $\id$ \\
\end{tabular}
\vskip 2mm \caption{Fusion rules of the parafermion and spin
fields associated to the parafermion theory SU$(3)_2/[$U(1)$]^2$
introduced by Gepner \protect\cite{gep1}.} \label{fusrul}
\end{table}

\begin{table}
\begin{tabular}{c|c|c}
& $\Delta N_\phi = 1$ & $\Delta N_\phi = 2$ \\ \hline
$N = 4$ &
$ \begin{array}{l|rcc}
\# =20 & S=0 & 1 & 2 \\
\hline
L=0 & 1 & 0 & 1 \\
L=1 & 0 & 1 & 0 \\
L=2 & 1 & 0 & 0 \\
\end{array}$ &
$\begin{array}{l|rcc}
\# =104 & S=0 & 1 & 2 \\
\hline
L=0 & 1 & 0 & 1 \\
L=1 & 0 & 2 & 0 \\
L=2 & 2 & 1 & 1 \\
L=3 & 0 & 1 & 0 \\
L=4 & 1 & 0 & 0 \\
\end{array}$ \\ \hline
$N = 8$ &
$\begin{array}{l|rcc}
\# =105 & S=0 & 1 & 2 \\
\hline
L=0 & 2 & 0 & 1 \\
L=1 & 0 & 2 & 0 \\
L=2 & 2 & 1 & 1 \\
L=3 & 0 & 1 & 0 \\
L=4 & 1 & 0 & 0 \\
\end{array}$ &
$\begin{array}{l|rcccc}
\# =1719 & S=0 & 1 & 2 & 3 & 4 \\
\hline
L=0 & 4 & 1 & 3 & 0 & 1 \\
L=1 & 1 & 7 & 2 & 1 & 0 \\
L=2 & 7 & 7 & 6 & 1 & 0 \\
L=3 & 3 & 9 & 3 & 1 & 0 \\
L=4 & 6 & 6 & 4 & 0 & 0 \\
L=5 & 2 & 5 & 1 & 0 & 0 \\
L=6 & 3 & 2 & 1 & 0 & 0 \\
L=7 & 0 & 1 & 0 & 0 & 0 \\
L=8 & 1 & 0 & 0 & 0 & 0 \\
\end{array}$\\ \hline
$N = 12$ &
$\begin{array}{l|rcc}
\# =336 & S=0 & 1& 2 \\
\hline
L=0 & 3 & 0 & 1 \\
L=1 & 0 & 3 & 0 \\
L=2 & 3 & 3 & 2 \\
L=3 & 1 & 3 & 0 \\
L=4 & 2 & 1 & 1 \\
L=5 & 0 & 1 & 0 \\
L=6 & 1 & 0 & 0 \\
\end{array}$ & \\ \hline \hline
&  $\Delta N_\phi = 3$ & $\Delta N_\phi = 4$ \\
\hline
$N = 4$ &
$ \begin{array}{l|rcc}
\# = 321 & S=0 & 1 & 2 \\
\hline
L=0 & 2 & 0 & 1 \\
L=1 & 0 & 2 & 0 \\
L=2 & 2 & 2 & 2 \\
L=3 & 1 & 3 & 0 \\
L=4 & 2 & 1 & 2 \\
L=5 & 0 & 1 & 0 \\
L=6 & 1 & 0 & 0 \\
\end{array}$ &
$\begin{array}{l|rcc}
\# = 755 & S=0 & 1 & 2 \\
\hline
L=0 & 2 & 0 & 1 \\
L=1 & 0 & 3 & 0 \\
L=2 & 3 & 2 & 2 \\
L=3 & 1 & 4 & 1 \\
L=4 & 3 & 3 & 2 \\
L=5 & 1 & 3 & 0 \\
L=6 & 2 & 1 & 1 \\
L=7 & 0 & 1 & 0 \\
L=8 & 1 & 0 & 0 \\
\end{array}$ \\
\end{tabular}
\vskip 2mm
\caption{Counting results for the NASS states at $k=2$. $N$ is the
number of electrons; $\Delta N_\phi$ is the number of excess flux
quanta. The results are given as a function of the $L$
(angular momentum) and $S$ (total spin) quantum numbers. The total
number of states is also indicated.}
\label{count1}
\end{table}

\begin{table}
\begin{tabular}{c|c|c}
& \rule[-2 mm]{0 mm}{6 mm}
$\Delta N_\phi = \frac{1}{2}$ & $\Delta N_\phi = \frac{3}{2}$ \\
\hline
$N = 2$ &
$ \begin{array}{l|rc}
\# =3 & S=0 & 1 \\
\hline
L=0 & 0 & 1 \\
\end{array}$ &
$\begin{array}{l|rc}
\# =10 & S=0 & 1 \\
\hline
L=0 & 1 & 0 \\
L=1 & 0 & 1 \\
\end{array}$ \\ \hline
$N=6$ &
$ \begin{array}{l|rc}
\# = 10  & S=0 & 1 \\
\hline
L=0 & 1 & 0 \\
L=1 & 0 & 1 \\
\end{array}$ &
$\begin{array}{l|rccc}
\# = 175 & S=0 & 1 & 2 & 3 \\
\hline
L=0 & 0 & 2 & 0 & 1 \\
L=1 & 2 & 1 & 1 & 0 \\
L=2 & 0 & 3 & 1 & 0 \\
L=3 & 2 & 1 & 0 & 0 \\
L=4 & 0 & 1 & 0 & 0 \\
\end{array}$ \\ \hline \hline
& \rule[-2 mm]{0 mm}{6 mm}
$\Delta N_\phi = \frac{1}{4} $ &
$\Delta N_\phi = \frac{5}{4} $ \\ \hline
$N = 5$ &
$ \begin{array}{l}
\# = 0  \\
\end{array}$ &
$ \begin{array}{l|rc}
\# = 48 \rule[-2 mm]{0 mm}{6 mm} & S= \frac{1}{2} & \frac{3}{2} \\
\hline
\rule[-2 mm]{0 mm}{6 mm} L= \frac{1}{2} & 1 & 1 \\
\rule[-2 mm]{0 mm}{6 mm} L= \frac{3}{2} & 1 & 1 \\
\rule[-2 mm]{0 mm}{6 mm} L= \frac{5}{2} & 1 & 0 \\
\end{array}$ \\ \hline \hline
& \rule[-2 mm]{0 mm}{6 mm}
$\Delta N_\phi = \frac{3}{4}$ &
$\Delta N_\phi = \frac{7}{4}$ \\ \hline
$N = 3$ &
$ \begin{array}{l|r}
\# = 4 \rule[-2 mm]{0 mm}{6 mm} & S= \frac{1}{2} \\
\hline
\rule[-2 mm]{0 mm}{6 mm} L= \frac{1}{2} & 1 \\
\end{array}$ &
$ \begin{array}{l|rc}
\# = 28 \rule[-2 mm]{0 mm}{6 mm} & S= \frac{1}{2} & \frac{3}{2} \\
\hline
L= 0 & 0 & 0 \\
L= 1 & 1 & 1 \\
L= 2 & 1 & 0 \\
\end{array} $ \\
\end{tabular}
\vskip 2mm
\caption{Counting results for the NASS states at $k=2$ with
fractional $\Delta N_\phi$ (symbols as in \protect\ref{count1}).}
\label{count2}
\end{table}

\end{multicols}
\end{document}